# Controlling nitrogen incorporation in (Eu,Ba,Ca)$TiO_3$-derived perovskite-type oxynitrides through Fermi Level Engineering

Margarida Barroso[1+], Marie Neumann[2+], Rachele Sciotto[3+], Shama Perween[4], Magdalena O. Cichocka[5], Jefferson Bettini[6], Adrián López Vergara[7], Patrick Zeller[2], Gaurav Anand[8], Verena Streibel[9], Barbara A. J. Lechner[8], Melanie Appelmann[10], Oliver Clemens[4], Ute Kolb[5], Jan P. Hofmann[2], Karsten Albe[3], Andreas Klein[7], Marc Widenmeyer[1*], Anke Weidenkaff[1]

[1] Technical University of Darmstadt, Department of Materials and Geosciences, Materials and Resources Laboratory, Peter-Grünberg-Straße 2, 64287 Darmstadt, Germany

[2] Technical University of Darmstadt, Department of Materials and Geosciences, Surface Science Laboratory, Peter-Grünberg-Straße 4, 64287 Darmstadt, Germany

[3] Technical University of Darmstadt, Department of Materials and Geosciences, Materials Modelling Division, Otto-Berndt-Straße 3, 64287 Darmstadt, Germany

[4] University of Stuttgart, Institute for Materials Science, Department of Chemical Materials Synthesis, Heisenbergstraße 3, 70569 Stuttgart, Germany

[5] Technical University of Darmstadt, Department of Materials and Geosciences, Electron Crystallography Laboratory, Schnittspahnstraße 9, 64287 Darmstadt, Germany

[6] Brazilian Nanotechnology National Laboratory (LNNano), Brazilian Center for Research in Energy and Materials (CNPEM), 13083-970 Campinas, São Paulo, Brazil

[7] Technical University of Darmstadt, Department of Materials and Geosciences, Electronic Structure of Materials, Otto-Berndt-Straße 3, 64287 Darmstadt, Germany

[8] Technical University of Munich, TUM School of Natural Sciences, Department of Chemistry, Functional Nanomaterials Group and Catalysis Research Center, Lichtenbergstr. 4, 85748 Garching, Germany

[9] Technical University of Munich, Walter Schottky Institute and TUM School of Natural Sciences, Physics Department, Am Coulombwall 4, 85748 Garching, Germany

[10] Johannes Gutenberg University Mainz, Centre for High Resolution Electron Microscopy (EMC-M), Duesbergweg 10-14, 55128, Mainz, Germany

[+] These authors contributed equally to this work

* corresponding author

## Abstract

Defect states directly influence not only functional properties, but also oxynitride formation during ammonolysis. Along with structural stability and charge balance, these defects can cause non-uniform nitrogen incorporation. To address this, the recently proposed Fermi Level Engineering concept provides a powerful alternative of rationalising oxynitride synthesis. By using the electron chemical potential, i.e., the Fermi level, as a unique descriptor, this approach can predict, understand, and manipulate defect formation enthalpies and charge transition levels, subsequently allowing to incorporate defect-property correlations to the concept. This work combines theoretical modelling and experimental characterisation of barium titanium oxynitrides to explain an intrinsic, self-limiting nitridation process. Progressive nitrogen incorporation causes a Fermi level downward shift, leaving further nitrogen substitution on oxygen sites energetically unfavourable, up to 0.03 N per unit formula. We demonstrate that *A*-site cation substitution in $Eu_{1-x}Ba_xTiO_3$- and $Eu_{1-x}Ca_xTiO_3$-derived oxynitrides can be applied to control compensation of the additional negative charge from $N^{3-}$ and keep the Fermi level high. The $Eu^{2+/3+}$ redox couple enables substitution of oxygen by nitrogen through pinning of the Fermi level at the europium charge transition level, at ~ 2.9 eV above the VBM. This allows for higher nitrogen uptake of up to 0.65 per formula unit (taken from hot gas extraction) through a more controlled process. Further incorporation is then limited by an intra-$Ti(O,N)_6$ octahedra distortion due to internal bond-angle variance. These findings demonstrate that Fermi Level Engineering provides a pathway to predict and overcome intrinsic synthetic and thermodynamic limitations in oxynitride synthesis.

## Introduction

Oxynitride materials attract considerable attention given their ability to combine the structural robustness of oxides with the favourable electronic properties introduced by nitrogen. In most cases, substituting $O^{2-}$ by $N^{3-}$ raises the valence band maximum due to the higher-energy N 2*p* states, narrowing the bandgap and improving visible light absorption compared to oxide materials [1,2]. An exception are Eu-containing titanates, that exhibit a widening of the bandgap after ammonolysis due to the involvement of the Eu 4*f* states [3]. Perovskite-type oxynitrides are especially promising due to versatile cation substitutions that permit tuning of the optical, electronic, and defect-related properties [4]. They are therefore interesting for solar water splitting (SWS), which requires efficient light absorption, suitable bandgaps, and favourable band edge positions for the water redox reactions [4–6]. Most bulk or powdered oxynitrides are currently synthesised by thermal ammonolysis of oxide precursors under flowing ammonia gas at high temperatures [7,8], and the compositional choice is guided typically by crystallographic stability and charge compensation strategies derived from the final products. However, ammonolysis is heavily influenced by topotactical reaction kinetics, often resulting in non-uniform nitrogen incorporation and a high concentration of defects such as anion vacancies/deficiencies and reduced transition-metal cations such as $Ti^{3+}$ [5]. This reduction and consequent charged defects affect the electronic structure, e.g., by moving the Fermi level ($E_F$) and increasing charge carrier recombination, which can significantly hinder the photocatalytic activity even when the bandgap seems suitable [5].

Cation substitution has served often as a tool for charge compensation and structure stabilisation, rather than as an intentional strategy to control defect equilibria. Recently, a more targeted (co-)substitution methodology for better control of oxynitride formation was introduced: the concept of Fermi Level Engineering (FLE) [9]. The Fermi level represents the electron chemical potential and is directly related to the defect concentration in a material and its defect chemistry. FLE takes the Fermi level as a common descriptor for the different charge compensation mechanisms that occur during synthesis and material treatments. This generalisation is possible based on the interdependence of defect concentrations and charge neutrality, governed by the Fermi level position: $E_F$ is determined by the concentration of charged defects and their associated energy levels. In addition, defect concentrations themselves are also closely linked to the Fermi level, since their formation enthalpies vary as a function of the $E_F$ position within the band structure [9]. Hence, the formation enthalpies of charged point defects such as oxygen vacancies, cation vacancies, reduced transition-metal species, and nitrogen dopants all depend on the Fermi level. Therefore, the $E_F$ directly governs which charge compensation mechanisms are thermodynamically favoured. During ammonolysis of oxides, for instance, acceptor doping by $N^{3-}$ downshifts the Fermi level, lowering the formation enthalpy of charge compensating donor defects such as oxygen vacancies and reduced cations. In turn, these defects can limit the downward movement of the $E_F$. These shifts influence the incorporation of nitrogen in the anionic sublattice. Charge transition levels (CTLs) are the values of the electron chemical potential at which defects shift charge states. By mapping CTLs and adjusting the formation enthalpies of relevant defects through different synthesis conditions and substitution [9], Fermi Level Engineering offers a unifying, predictive concept that connects synthesis conditions, defect thermodynamics, and functional properties, which can be applied to oxynitride materials as photoabsorbers.

Within a FLE perspective, $BaTiO_3$ (BTO) serves as an ideal model system. It maintains the perovskite-type structure over wide temperature ranges, benefits from extensive available literature on defect properties, and shows commercial and technological relevance [10–12]. Upon ammonolysis of pristine $BaTiO_3$, the available charge compensation mechanism for $N^{3-}$ acceptor doping is the formation of oxygen vacancies. Perovskite-type materials generally remain stable against transformation into lower-coordination phases (e.g., brownmillerite) up to an anion-deficiency of $\delta \approx 0.5$ [13]. Theoretically, this stability window would allow up to one nitride ion per unit formula, leading to $BaTiO_{3-3\delta}N_{2\delta}$ ($0 < \delta < 0.5$). In practice, however, reported N contents in pristine $BaTiO_3$ are below 0.1 N per formula unit [14,15]. While higher nitrogen contents (up to 0.4) have been observed, they require different starting materials, for instance, reduced $BaTi(O_{3-x}H_x)$ oxyhydride [16,17] or Ta-substituted $BaTiO_3$ [18].
In this study, direct ammonolysis of pristine $BaTiO_3$ was deliberately chosen to keep a single-step synthesis that avoided intermediates, co-dopants, or additional chemical processes. This choice could minimise the environmental footprint which is typical of oxynitride synthesis routes [19], while preserving a simple baseline system. From an FLE standpoint, nitrogen incorporation in pristine BTO might drive excessive $E_F$ downward shifts, possibly beyond favourable energetic values. Introducing heterovalent cations such as Eu on the *A*-site bypasses such excessive shifts by introducing another charge compensation mechanism. In $EuTiO_3$, which preserves the perovskite-type structure across both the oxide and resulting oxynitride configuration, the $Eu^{2+}/Ti^{4+}$ base state allows oxidation of $Eu^{2+}$ to $Eu^{3+}$ to balance $N^{3-}$ uptake [3,20]. This modifies defect formation enthalpies and drives the $E_F$ stabilisation mechanism away from being strictly vacancy-based. Furthermore, to allow even further nitrogen incorporation, isovalent substitution of $Ba^{2+}$ for smaller $Ca^{2+}$ changes lattice strain and electronic structure, thereby influencing the density of states of the Eu 4*f* levels [21,22].

To bridge these gaps and investigate the thermodynamic limitations preventing stoichiometric nitrogen incorporation, this work combines Density Functional Theory (DFT) modelling with experimental synthesis of $BaTiO_3$ and its oxynitride to map defect energetics and Fermi level shifts. Building on this reference system, we extend our investigation to $Eu_{1-x}Ba_xTiO_3$ and $Eu_{1-x}Ca_xTiO_3$ and corresponding oxynitrides to deliberately manipulate the $E_F$ position through the $Eu^{2+/3+}$ charge transition level. By using complementary structural and spectroscopic tools such as neutron diffraction, hot gas extraction (HGE), UV Visible Spectroscopy, X-Ray Photoelectron Spectroscopy (XPS) and reductive and oxidative *in situ* treatments, we established a direct correlation between defect chemistry, crystal and electronic structure, and nitrogen incorporation. This correlation paves the way towards a more complete design of functional oxynitrides through simultaneous engineering of their structural and electronic properties.

## Results and Discussion

### 1. Ammonolysis of pristine $BaTiO_3$

#### *i. Effect of defect formation energies on incorporation of nitrogen*

DFT calculations were performed on rhombohedral $BaTiO_3$ using the Python package pynter [23]. The defect model includes oxygen vacancies $v_O$ with charge states $q = 0, +1, +2$, and nitrogen substitution on the oxygen sites $N_O$ with charge states $q = -2, -1, 0, +1$. Here, $q$ denotes the net defect charge and should not be interpreted as the formal oxidation state of N. Defect formation energies were calculated according to Equation S1, using the chemical potentials corresponding to point X in the stability diagram (Figure S1.1). Charged-defects finite-size corrections were applied using the scheme proposed by Kumagai *et al.* [24]. The relevant charge states and charge transition levels (CTLs, Equation S3) were included in the charge-neutrality condition together with electrons and holes. Electronic carrier concentrations were determined from the bulk density of states. Under the conditions considered here, the hole concentration is negligible.

The data analysis was performed with the Python package defermi [25]. We first investigate the electronic response to nitrogen incorporation by treating the total nitrogen concentration [N] as an externally imposed quantity. For each imposed [N], the charge-neutrality condition was solved self-consistently for the Fermi level. Increasing [N] shifts $E_F$ toward lower energies (Figure 1a), suggesting the feedback mechanism illustrated in Figure 1b: nitrogen incorporation lowers $E_F$, which in turn affects the thermodynamic driving force for further $N_O$ formation.

This behaviour can be explained by the defect formation energies in Figure 1c, evaluated at fixed chemical potentials corresponding to the selected experimental conditions and plotted with transposed axes. The horizontal dashed lines in Figure 1a and 1c indicate two representative Fermi level positions (2.95 and 2.55 eV above the valence-band maximum) and serve as guides to the eye, illustrating how the shift in $E_F$ affects the formation energies of $N_O$ and $v_O$. The stars represent the CTLs, while the crossing point between the formation energies of $N_O$ and $v_O$ marks the Fermi level at which the two defects become energetically equivalent. Above this crossing, $N_O$ has the lower formation energy, whereas below it, $v_O$ becomes energetically more favourable. As the imposed nitrogen concentration increases (Figure 1a), the Fermi level shifts toward lower energies, moving further below the crossing point. This increases the formation energy of $N_O$, making further nitrogen incorporation less favourable. Figure 1a describes the response to an imposed nitrogen concentration rather than the equilibrium nitrogen uptake at fixed nitrogen partial pressure.

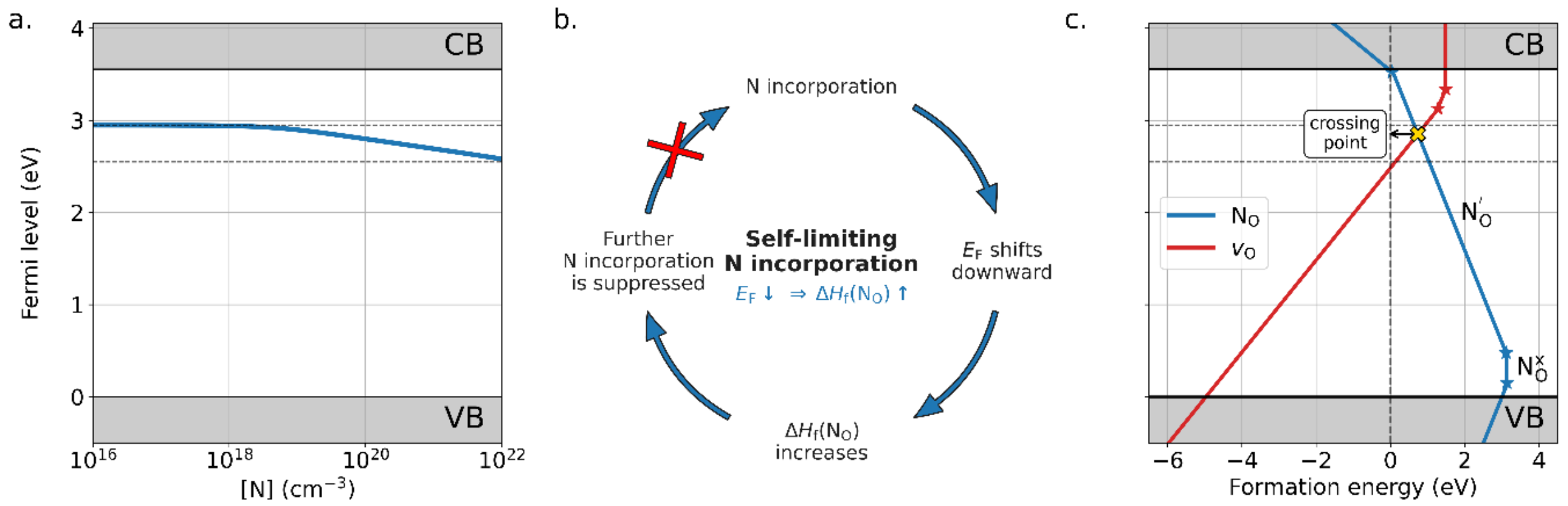

Figure 1. a. Self-consistent Fermi level $E_F$ obtained from the charge-neutrality condition as a function of imposed nitrogen concentration [N] at $T$ = 1073 K. b. Schematic representation of the proposed self-limiting feedback mechanism: nitrogen incorporation lowers $E_F$, thereby increasing the formation energy of $N_O$ and suppressing further nitrogen incorporation. c.

Formation energies of $N_O$ and $v_O$ as a function of $E_F$, plotted with transposed axes. The horizontal dashed lines in Figure 1a. and 1c. mark two Fermi level positions and serve as guides to the eye for visualizing the effect of the Fermi-level shift on defect formation energies. Stars indicate the corresponding charge transition levels, and the marked crossing point indicates equal formation energies of $N_O$ and $v_O$. Formation energies were determined at $T$ = 1073 K and $p_{O_2} = 10^{-27}$ atm, with the nitrogen chemical potential corresponding to an ideal-gas $N_2$ reservoir with $p_{N_2}$ = 0.25 atm, estimated assuming complete $NH_3$ dissociation at a total pressure of 1 atm.

The model therefore identifies an electronic negative-feedback mechanism that can contribute to self-limiting nitridation: incorporation of $N_O$ lowers the Fermi level, which raises the formation energy of nitrogen on oxygen and reduces the driving force for further incorporation.

### *ii. Crystal and electronic structure invariance in BaTi(O,N)$_3$ vs BaTiO$_3$*

Thermal ammonolysis of $BaTiO_3$ was performed. The refinements of the XRD data of the oxide and resulting oxynitride can be seen in Figure 2. As expected, the diffractogram remains essentially unchanged from that of the parent oxide, in the tetragonal *P*4*mm* space group, and the lattice parameters remain very similar. The refinements of BaTi(O,N)$_3$ prepared at $T$ = 973 K and further details are found in Figure S2.1 and Table S2.1.

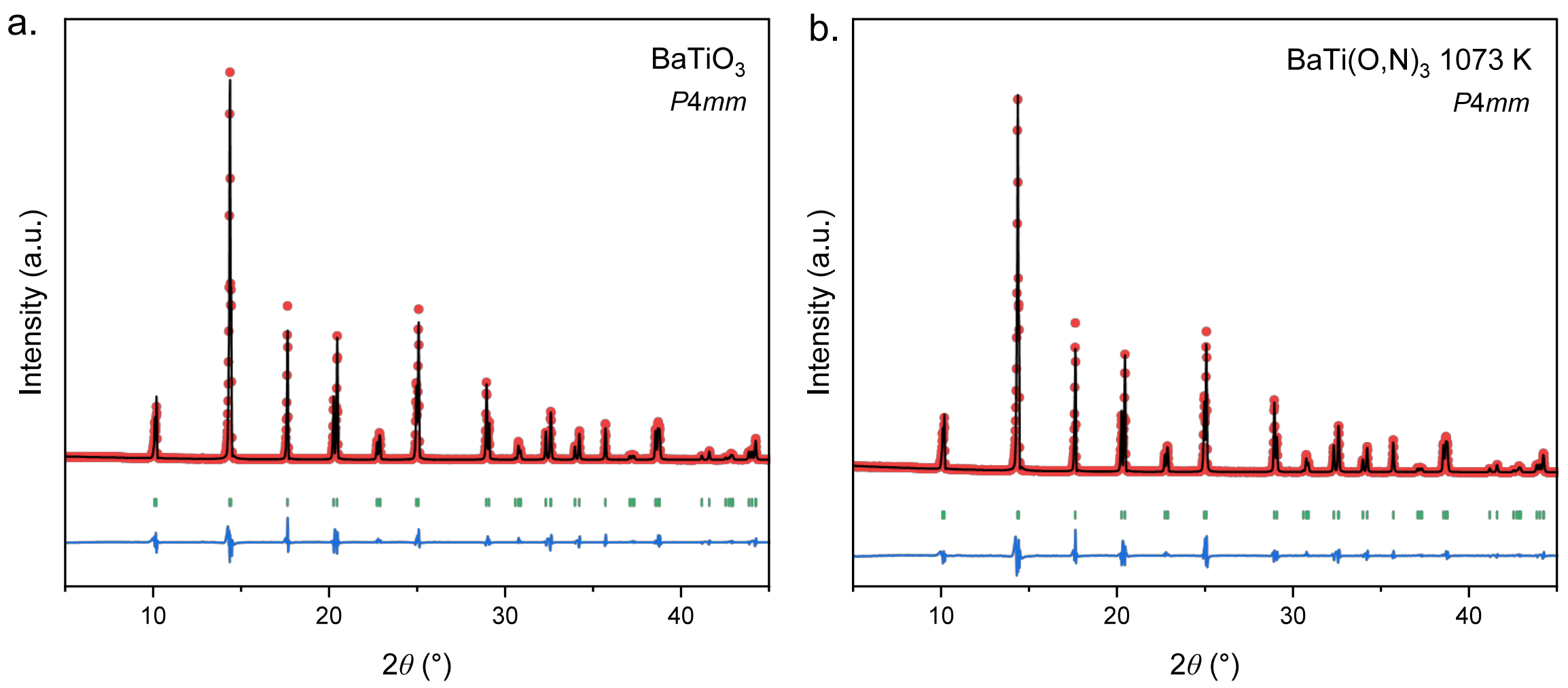

Figure 2. Refined XRD ($\lambda$ = 0.70932 Å) patterns of a. $BaTiO_3$, and b. BaTi(O,N)$_3$ synthesised at $T$ = 1073 K, using the tetragonal *P*4*mm* space group model. The red dots represent the experimental data, the black line the calculated pattern, and the blue line the difference curve. The green hash marks represent the indexed reflections of the model.

The oxygen and nitrogen contents, displayed in Table 1, demonstrate that the nitrogen incorporation in pristine $BaTiO_3$ is exceptionally low, at a maximum of N ≈ 0.033 atoms per formula unit. This is lower than values observed in literature [15,26], and confirms the predictions made by density functional theory.

Table 1. Oxygen and nitrogen contents of $BaTiO_3$ and resulting oxynitrides, determined by hot gas extraction (HGE).

| Sample | Composition |
|---|---|
| $BaTiO_3$ | $BaTiO_{3.06\pm0.05}$ |
| BaTi(O,N)$_3$ (973 K) | $BaTiO_{2.99\pm0.02}N_{0.015\pm0.003}$ |
| BaTi(O,N)$_3$ (1073 K) | $BaTiO_{2.967\pm0.001}N_{0.033\pm0.001}$ |

The optical properties of the samples were measured through UV-Visible spectroscopy (Figure S3.1). As expected, the optical bandgap of both oxynitrides remains identical to the bandgap of $BaTiO_3$, with $E_g$ = 3.2 eV for the parent oxide and both oxynitrides. This provides further evidence that the introduction of nitrogen is very low, keeping the electronic band structure essentially unchanged. A smaller feature can be observed at 2.5 eV. However, rather than representing a true band-to-band transition, this sub-bandgap absorption tail is likely induced by isolated, in-gap N 2*p* states situated just above the VB.

XPS was performed (results seen in Figure S4.1); however, given the low nitrogen content in the ammonolysed materials, the system maintains the electronic characteristics of $BaTiO_3$, which has very poor electrical conductivity. Consequently, significant charging effects were observed during the XPS measurements. Because an electron flood gun was used to compensate for surface charging, the resulting shifts and manual treatment of the data compromised the determination of the $E_F$.

Ultimately, the low nitrogen incorporation observed in pristine $BaTiO_3$ is dictated by defect thermodynamics. Our DFT calculations reveal that the substitution of oxygen by nitrogen ($\mathrm{N_O}$) becomes energetically favourable only at high Fermi levels. Keeping the $E_F$ high is therefore a defining factor to allow higher nitrogen incorporation.

## 2. Cationic substitution improves nitrogen incorporation: $Eu_{1-x}Ba_xTi(O,N)_3$ and $Eu_{1-x}Ca_xTi(O,N)_3$ ($x$ = 0.2, 0.8)

Guided by the predictions from the electrochemical series for solid materials [27], we performed *A*-site substitution with europium to maintain the Fermi level high, closer to the CBM. We use the $Eu^{2+/3+}$ redox pair to ideally keep defect formation energies favourable for higher nitrogen incorporation. In addition, $Ca^{2+}$ substitution was applied on the *A*-site to induce lattice strain, serving as a complementary pathway to promote higher nitrogen contents [3].

Accurately describing the influence of the localised Eu 4*f* states is challenging for semi-local density functional approximations [28,29]. Similar difficulties are encountered by DFT+U or hybrid-functional approaches [30,31]. Therefore, within the PBE-level framework used in this work, we restrict the quantitative charge-transition-level analysis to $\mathrm{N_O}$ and $v_\mathrm{O}$.

### *i. Facilitation of nitrogen uptake by lower symmetry*

The Powder Neutron Diffraction (PND) data refinements, shown in Figure 3, confirm that $Eu_{0.2}Ba_{0.8}Ti(O,N)_3$ crystallises in a *P*4*mm* tetragonal structure, maintaining the symmetry of the parent oxide, as shown in Figure S5.1. It contains a minor secondary phase of titanium nitride (space group $Fm\bar{3}m$). This is not unexpected, as it is a classic by-product of the strongly reducing ammonolysis environment, leading to local reduction of Ti [3].

$Eu_{0.8}Ba_{0.2}Ti(O,N)_3$, however, undergoes loss of translational symmetry, with superstructure reflections indicative of the orthorhombic space group *Pnma*. The multi-cubic-phased parent oxide (Figure S5.1), $Eu_{0.8}Ba_{0.2}TiO_3$, likely contains localised microstrain gradients and compositional inhomogeneity [32], which are likely to be driven by the presence of mixed-valence states ($Eu^{2+}/Eu^{3+}$). Remarkably, no indication for such inhomogeneity from reflection profiles was indicated for the oxynitride phase.

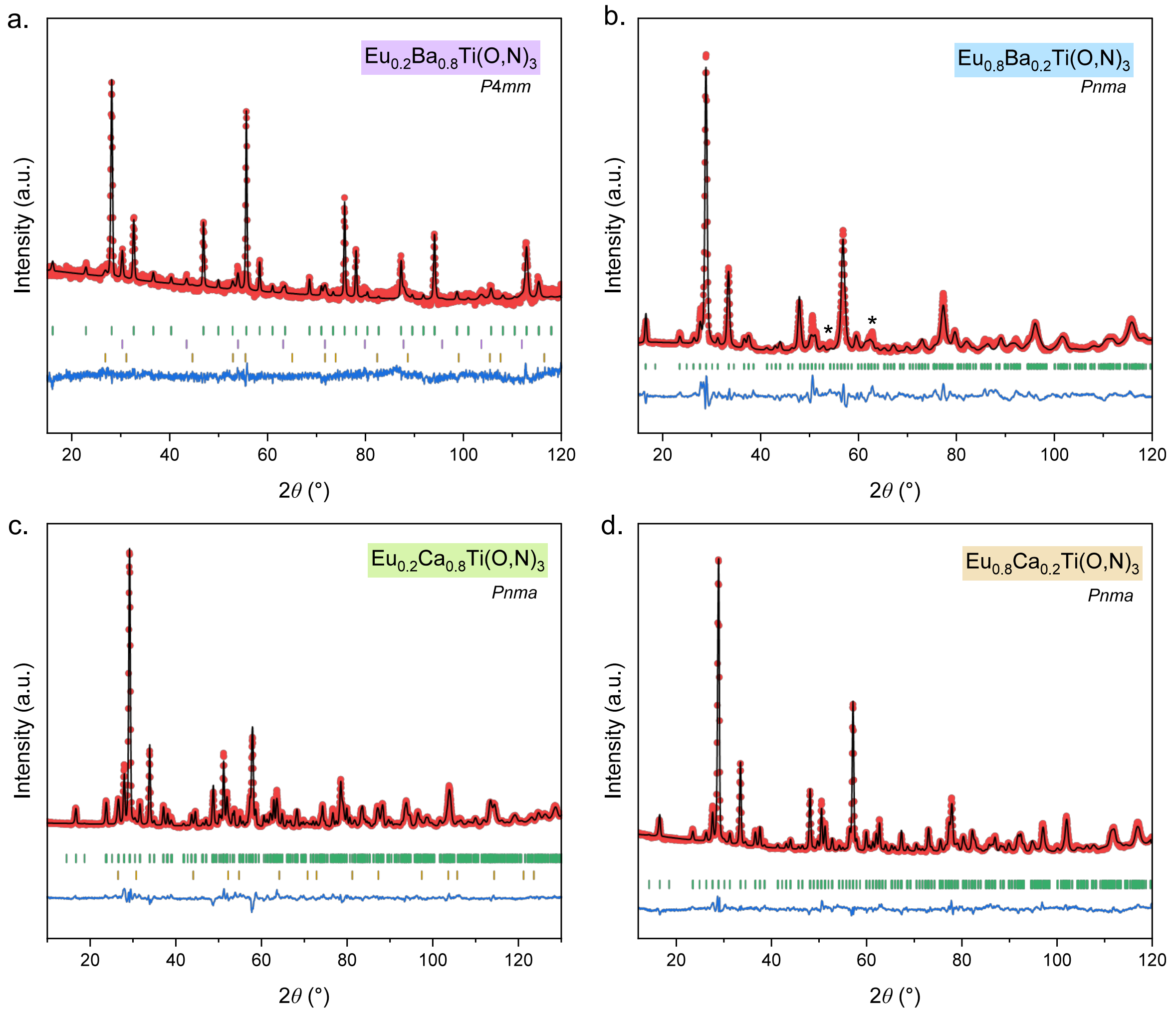

Figure 3. Refined neutron diffraction ($\lambda$ = 1.120148(1) Å) patterns of a. $Eu_{0.2}Ba_{0.8}Ti(O,N)_3$, using the tetragonal *P4mm* space group model; and b. $Eu_{0.8}Ba_{0.2}Ti(O,N)_3$, c. $Eu_{0.2}Ca_{0.8}Ti(O,N)_3$, and d. $Eu_{0.8}Ca_{0.2}Ti(O,N)_3$, using the orthorhombic *Pnma* space group model. The red dots represent the experimental data, the black line the calculated pattern, and the blue line the difference curve. The green hash marks represent the indexed reflections of the respective *P4mm* and *Pnma* models, the purple ones the vanadium can reflections, and the yellow ones the reflections of TiN. The * also represents vanadium can reflections.

From the refinements of the Ba-containing oxides, the Eu:Ba ratios were 0.216(1):0.784(1) and 0.810(1):0.190(1). The refined O/N occupancies on the resulting oxynitrides led to $Eu_{0.2}Ba_{0.8}TiO_{2.528(1)}N_{0.184(1)}\square_{0.288(1)}$ and $Eu_{0.8}Ba_{0.2}TiO_{2.8(1)}N_{0.1(1)}\square_{0.0(1)}$.

The PND confirms that both Ca-containing oxynitrides show an orthorhombic *Pnma* perovskite-type structure. While the Eu-poor sample mainly maintains the structure from its parent oxide (Figure S5.2), $Eu_{0.8}Ca_{0.2}Ti(O,N)_3$ undergoes slight structural transition. $Eu_{0.8}Ca_{0.2}TiO_3$ (Figure S5.2) was modelled using the *Pnma* space group, as a tetragonal symmetry alone was insufficient to refine it. However, the pattern shows lack of certain reflections predicted by the orthorhombic symmetry. This might arise due to a local-scale competition between *Pnma* and *I*4/*mcm* domains, similar to structural fluctuations previously reported [33,34]. While this Eu-rich oxynitride shows no secondary phases in PND, for $Eu_{0.2}Ca_{0.8}Ti(O,N)_3$, there is a small impurity phase of TiN, suggesting minor decomposition during the ammonolysis, which is in accordance with previous findings [3]. Further 3D Electron Diffraction (3DED) analysis of $Eu_{0.2}Ca_{0.8}Ti(O,N)_3$ revealed, in addition to the Ca-rich *Pnma* majority phase previously obtained from PND and HR synchrotron XRD, a minority cubic phase with $P2_13$ symmetry. Dynamical refinements of the *Pnma* phase ($R$ = 13.80 %) gives an Eu:Ca ratio of 17:83, in good agreement with the bulk composition (20:80), whereas dynamical refinement of the $P2_13$ phase ($R$ = 25.99 %) yields nearly equimolar Eu/Ca occupancies, demonstrating its compositionally distinct character. Additionally, the minority phase inclusion in the structural model accounts for a series of very weak reflections close to the detection limit in the HR synchrotron pattern of $Eu_{0.2}Ca_{0.8}TiO_3$ indicating that the $P2_13$ phase is likely an oxide. The crystallographic details derived from 3DED are summarized in Table S5.10. Complementary Electron Energy-Loss Spectroscopy (EELS) measurements reveal Ti-rich nanoscale domains, consistent with the presence of TiN at the nanoscale (Figure S6.1).

From the precursor oxides high resolution XRD data (Tables S5.7., S5.8., and S5.9.), the refined ratios of Eu:Ca led to 0.192(1):0.808(1) and 0.796(2):0.204(2). These are close to the results obtained by SEM-EDX, which show Eu:Ca ratios of 0.23(2):0.82(3) and 0.79(4):0.21(2). The refinements of the O/N ratios led to $Eu_{0.2}Ca_{0.8}TiO_{2.55(2)}N_{0.36(2)}\square_{0.09(2)}$ and $Eu_{0.8}Ca_{0.2}TiO_{2.20(4)}N_{0.59(4)}\square_{0.21(4)}$, similar to the results obtained by hot gas extraction.

In all our oxynitrides, N showed preference for the axial 4*c* sites compared to the equatorial 8*d* sites, indicating partial site-selective anion ordering. While many oxynitrides favour equatorial positioning via local *cis*-octahedral configurations [35], there is indication in literature of the possibility of some *B* = Ti oxynitrides favouring a local *trans*-configuration [36].

The oxygen and nitrogen contents were measured by hot gas extraction and can be seen in Table 2, where a trend of less to most N content is revealed: $Ba_{0.8}$ < $Ca_{0.8}$ < $Ba_{0.2}$ < $Ca_{0.2}$.

Table 2. Oxygen and nitrogen contents measured by HGE of $Eu_{1-x}Ba_xTi(O,N)_3$ and $Eu_{1-x}Ca_xTi(O,N)_3$ ($x$ = 0.2, 0.8).

| Sample | Composition |
|---|---|
| $Eu_{0.2}Ba_{0.8}Ti(O,N)_3$ | $Eu_{0.2}Ba_{0.8}TiO_{2.63\pm0.02}N_{0.14\pm0.01}\square_{0.23\pm0.02}$ |
| $Eu_{0.8}Ba_{0.2}Ti(O,N)_3$ | $Eu_{0.8}Ba_{0.2}TiO_{2.34\pm0.07}N_{0.44\pm0.01}\square_{0.22\pm0.07}$ |
| $Eu_{0.2}Ca_{0.8}Ti(O,N)_3$ | $Eu_{0.2}Ca_{0.8}TiO_{2.78\pm0.04}N_{0.21\pm0.01}\square_{0.01\pm0.04}$ |
| $Eu_{0.8}Ca_{0.2}Ti(O,N)_3$ | $Eu_{0.8}Ca_{0.2}TiO_{2.21\pm0.01}N_{0.65\pm0.01}\square_{0.14\pm0.01}$ |

While pure $BaTiO_3$ resists nitridation, with [N] < 0.1 per formula unit, the introduction of europium acts as an electronic buffer that enables higher nitrogen content across all samples. Given that europium is mainly in the 3+ state in the oxynitrides (as discussed in Section 2.ii. ahead), the defect chemistry is consistent with $Eu^{3+}/N^{3-}$ charge compensation.

Despite the shared electronic stabilisation, the calcium series achieves higher nitrogen uptake than the barium series due to differences in lattice flexibility and local distortion. For $Eu_{1-x}Ba_xTi(O,N)_3$, at high Ba content the larger ionic radius of $Ba^{2+}$ ($r_i$ = 1.61 Å) with Eu substitution takes the lattice framework into a rigid, non-tilted tetragonal *P*4*mm* space group, as verified by neutron diffraction. Because the *P*4*mm* structure does not distort to relieve the strain, forcing $N^{3-}$ into this tightly packed lattice leads to structural limits at N ≈ 0.14 ($x$ = 0.8), favouring the formation of oxygen vacancies to stabilise the system. In contrast, at high Eu content the system relaxes into an orthorhombic structure (space group *Pnma*) allowing to relieve lattice strain by using the $a^+b^-b^-$ tilting pattern. However, the size mismatch between $Ba^{2+}$ and $Eu^{3+}$ induces a larger distortion of the $Ti(O,N)_6$ octahedra by ca. 38% compared to $Eu_{0.8}Ca_{0.2}Ti(O,N)_3$ (Table S5.4.), limiting the N content to about 0.44.

The calcium series makes use of the structural flexibility given by the lower symmetry, orthorhombic *Pnma* space group. The smaller radius of $Ca^{2+}$ ($r_i$ = 1.34 Å) allows for $TiO_6$ octahedral tilting, and this counteracts the strain introduced by the $N^{3-}$ ions. Consequently, the measured nitrogen content of $Eu_{0.2}Ca_{0.8}Ti(O,N)_3$ (N ≈ 0.21) closely matches the Eu concentration measured in SEM-EDX and refined by HR-XRD, indicating a stoichiometric incorporation of N, without the need for significant oxygen vacancy formation. While the Eu-rich sample would ideally lead to an approximate N content of 0.8, the measured nitrogen content of $Eu_{0.8}Ca_{0.2}Ti(O,N)_3$ is N = 0.65, slightly higher than what was previously observed [3]. Ultimately, the europium's charge transition level allows for more controlled nitrogen incorporation and higher N contents. However, the perovskite-type lattice reaches a nitrogen incorporation limit, arising from increasing lattice strain, anion-site instability, and the lower thermodynamic stability against reductive decomposition at high nitrogen concentrations.

As the europium content is increased from 0.2 to 0.8, the UV-Vis absorption edges (Figure S7.1) shift slightly towards lower energies, with the extrapolated absorption maximum around 1.3 eV from the first absorption edge. However, low reflectance from the dark powders distorts the curves, which leads to large absorption tails caused likely by in-gap states, such as the Eu 4*f* states near the valence band. Therefore, it is likely that the extrapolations of the UV-Vis spectra are measuring transitions starting from these occupied states up to the CB, rather than revealing the true fundamental bandgap from the VB.

### *ii. Importance of the charge transition level on higher nitrogen incorporation*

To investigate the electronic structure and chemical states of the samples, synchrotron-based XPS was conducted. An analysis of the Eu 4*d* core-level spectra before and after ammonolysis reveals the co-existence of both $Eu^{2+}$ and $Eu^{3+}$ valence states in the initial oxides (Figure 4). To enable the direct comparison between different Eu concentrations, the oxynitride spectra were scaled so that the intensity around 124 eV to 126 eV and 146 eV to 150 eV aligned with the oxide.

The spin-orbit split multiplets of $Eu^{3+}$ (Eu $4d_{5/2}$ and $4d_{3/2}$) are clearly distinguished at approximately 136 eV and 142 eV, respectively. The spectroscopic identification of $Eu^{2+}$ is more subtle; its $4d_{3/2}$ component appears as a distinct shoulder on the lower binding energy side of the $Eu^{3+}$ $4d_{5/2}$ peak (around 134 eV). However, the corresponding $Eu^{2+}$ Eu $4d_{5/2}$ primary peak is split up into a broad feature spanning from 127 eV to 132 eV. This distinct configuration aligns with previous studies, which state that the intensity of this feature scales with the $Eu^{2+}$ concentration and is fundamentally driven by complex final state effects [37,38]. Benefiting from the high resolution of the synchrotron source, a characteristic fine structure within this region was successfully resolved, as previously seen in literature [38,39]. For further analysis, this complex structure is explicitly assigned to the $Eu^{2+}$ electronic states.

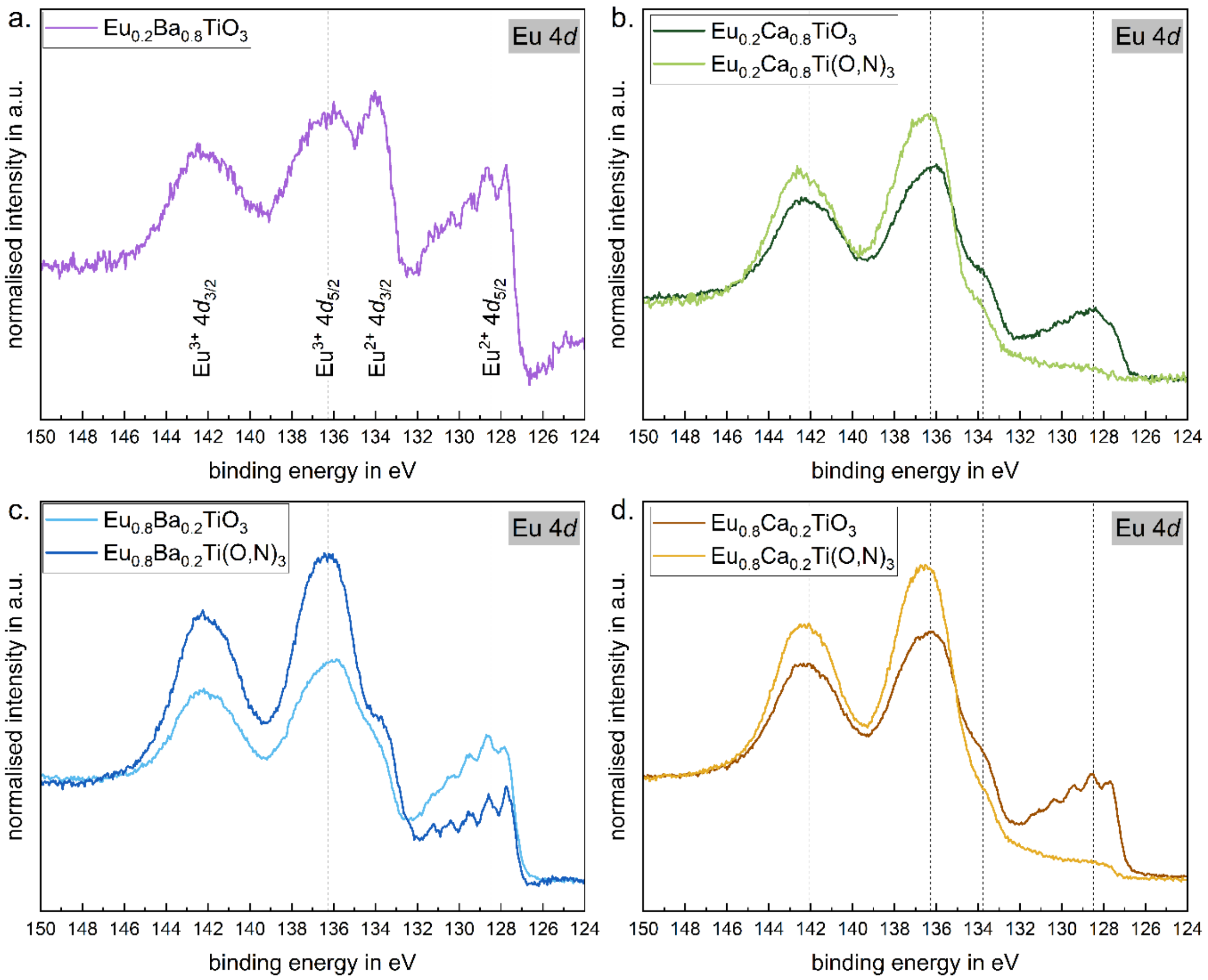

Figure 4. Synchrotron-based X-ray photoelectron spectra of the Eu 4*d* core levels for the perovskite-type oxides and their corresponding ammonolysed oxynitride counterparts, with a. and c. $Eu_{1-x}Ba_xTi(O,N)_3$, and b. and d. $Eu_{1-x}Ca_xTi(O,N)_3$. The spin-orbit split multiplets for $Eu^{3+}$ ($4d_{5/2}$ and $4d_{3/2}$) are indicated, alongside the characteristic low-binding-energy shoulder (~134 eV) and broad multi-peak feature (127–132 eV), associated with the screened final states of the $Eu^{2+}$ valence fraction. The oxynitride spectra were scaled so that the intensity around 146 eV to 150 eV aligned with the oxide. Measured at SoLiAS@OPUS at BESSY II with an excitation energy of $h\nu$ = 786 eV ($E_{kin}$ = ~650 eV).

Elemental ON analysis shows a clear trend in nitrogen incorporation: $Eu_{0.8}Ca_{0.2}$ > $Eu_{0.8}Ba_{0.2}$ > $Eu_{0.2}Ca_{0.8}$ > $Eu_{0.2}Ba_{0.8}$, while no major contribution of oxygen vacancies is measured. The Ti 2*p* synchrotron XPS data shows a related trend of titanium atoms in a mixed oxynitride coordination environment (see Figure S8.1 for the N 1*s* spectra, Figure S8.2 for the Ti 2*p* spectra). Due to the low nitrogen incorporation in the $Eu_{0.2}Ba_{0.8}Ti(O,N)_3$ sample, and it's low conductivity leading to significant surface charging, the $Eu_{0.2}Ba_{0.8}TiO_3$ and $Eu_{0.2}Ba_{0.8}Ti(O,N)_3$ samples were excluded during further discussion.

A comparison between the oxide and oxynitride spectra confirms that the $Eu^{2+}$ to $Eu^{3+}$ oxidation reaction is the primary charge compensation mechanism during ammonolysis [20]. Upon ammonolysis, $Eu^{2+}$ nearly completely converts into $Eu^{3+}$ for $Eu_{0.8}Ca_{0.2}TiO_3$. In contrast, $Eu_{0.8}Ba_{0.2}TiO_3$ shows an incomplete transition, retaining a significant amount of residual $Eu^{2+}$, as indicated by the persistently prominent feature around 127 eV to 132 eV and the lower N content. For the Eu-poor $Eu_{0.2}Ca_{0.8}TiO_3$ sample, while nearly all available $Eu^{2+}$ ions are successfully oxidised to $Eu^{3+}$, the lower europium concentration intrinsically limits the total capacity for nitrogen uptake.

Binding energy shifts were monitored by relating the shift of the Fermi level relative to the VBM, and comparing them to the shift of the O 1*s*, and Ti 2*p* core levels to track electronic variations using a methodology adapted from Chaoudhary et al. [40], where the onsets of the valence band spectra are directly read off from the measured valence band spectra, and the shifts of the core levels are indirectly compared by subtracting one fixed value for each core level, as shown in the graph (see Figure 5). This is done to evaluate whether the core level binding energy shift can be related to the Fermi level shift. To be able to do this, we need to ensure that all materials are sufficiently conductive (to mitigate charging effects) and that the valence band maximum does not change shape. The subtracted value was chosen by subtracting the VBM position from the O 1*s* or Ti 2*p* peak position of $Eu_{0.2}Ca_{0.8}TiO_3$ and extrapolating this to all other compounds to achieve a common scale. The Ti 2*p* spectra can be found in Figure S8.2, while the O 1*s* and VB spectra are compiled in Figure S8.3. The measured shifts can be seen in Table S8.1.

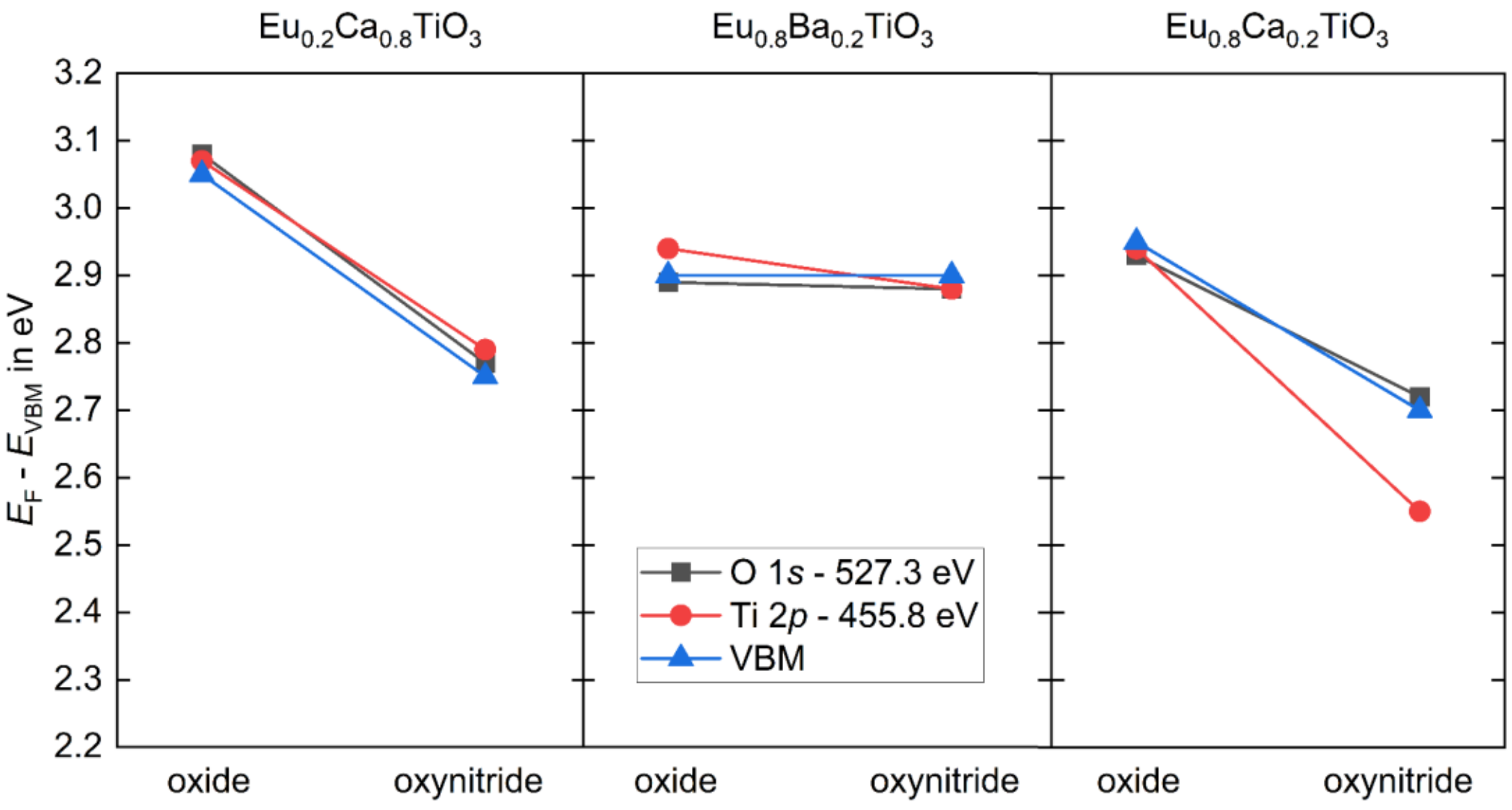

Figure 5. Fermi level shifts, determined from the valence band maximum (VBM) positions and indirectly from representative core levels (O 1*s*, Ti 2*p*), tracking the electronic variations upon ammonolysis, with errors of ± 0.1 eV. All determined core level values were subtracted by the values given in the legend. The lines connecting the oxide and oxynitride values are for visual guidance only. Measured at SoLiAS@OPUS at BESSY II, the detail spectra can be found in Figure S8.2 and Figure S8.3.

The Eu 4*d* level was excluded from this tracking due to its active participation in charge compensation, while N 1*s*, Ca 2*p*, and Ba 3*d* were omitted because they are not universally present across all samples. The survey scans of all samples, as well as the Ca 2*p*, Ba 3*d*, and C 1*s* spectra can be found in Figure S8.4 and Figure S8.5, and display similar shifts. Upon ammonolysis, a distinct downshift of the Fermi level to 2.75 eV (± 0.1 eV) above the VBM is observed for the Ca-containing samples, which is followed by the core levels. In contrast, the Fermi level of the Ba-containing system remains rigid within an experimental uncertainty of ± 0.1 eV. The incomplete $Eu^{2+}/Eu^{3+}$ transition in the Ba-containing oxynitride suggests that the Fermi level remains electronically pinned at the $Eu^{2+}/Eu^{3+}$ CTL, while the Ca-containing samples can complete the charge transition and unpin the Fermi level again. This locates the CTL of $Eu^{2+}/Eu^{3+}$ at around 2.9 eV (± 0.1 eV) above the VBM.

A significant discrepancy was detected in the Ti 2*p* binding energy position, specifically for the $Eu_{0.8}Ca_{0.2}Ti(O,N)_3$ sample. While all other elemental core levels moved in parallel, the anomalous downshift of the Ti 2*p* doublet by an additional 0.2 eV implies a distinct modification of the local coordination environment around the central titanium cation, which has been repeatedly observed in literature upon ammonolysis or substitution [2,3,20,41]. Further analysis of this shift was performed by fitting the Ti 2*p* spectra, as well as complementing the data with Near-Edge X-ray Absorption Fine Structure (NEXAFS) spectroscopy at the Ti L-edge (see Figure S8.2 and the additional discussion in S8.1).

To probe the dynamic behaviour of the electronic states and assess whether the Fermi level remains rigidly pinned or responds to external chemical potentials, a series of *in situ* oxidising and reducing thermal treatments was performed on the $Eu_{0.2}Ca_{0.8}Ti(O,N)_3$ oxynitride. The sample was heated to $T = 573$ K under alternating atmospheres of molecular oxygen ($5 \times 10^{-4}$ mbar $O_2$) or in ultra-high vacuum (UHV, $5 \times 10^{-9}$ mbar), and the Eu 4*d* and valence band spectra can be found in Figure 6.

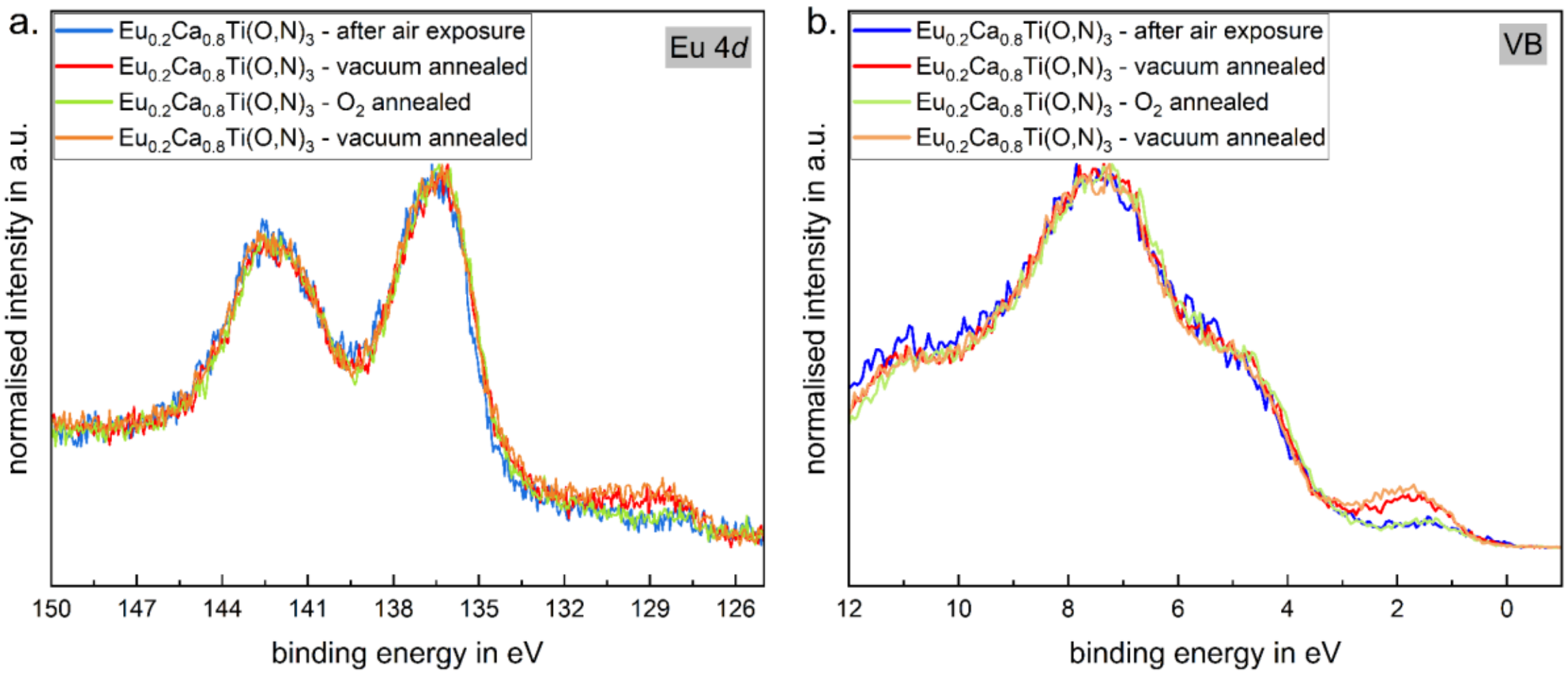

Figure 6. X-ray photoelectron spectra of a. the Eu 4*d* core levels and b. corresponding valence band for $Eu_{0.2}Ca_{0.8}Ti(O,N)_3$ during cyclic redox testing after annealing in vacuum ($1 \times 10^{-8}$ mbar during heating) or in oxygen partial pressure (backfilling the vacuum chamber with $5 \times 10^{-4}$ mbar $O_2$) at $T = 573$ K and cooling down to room temperature, measured at DAISY-FUN with Al $K_\alpha$ radiation. The spectra were normalised minimum to maximum.

Consistent with previous observations, the as-synthesized oxynitride surface is oxidised, with almost complete conversion of $Eu^{2+}$ to $Eu^{3+}$. Subsequent *in situ* annealing under reducing UHV conditions triggers a reduction of the surface, marked by a small intensity increase in both the $Eu^{2+}$ 4*d* final-state multiplet features (127–132 eV) and the characteristic $Eu^{2+}$-derived states located near the top of the valence band (≈ 2 eV below the Fermi level). Cyclic exposure to alternating oxidising and reducing atmospheres demonstrates that this low-temperature redox behaviour is fully reversible. No Fermi level shift is visible, suggesting either that, after the ammonolysis, the Fermi level is still pinned at the CTL of $Eu^{2+}$/ $Eu^{3+}$ (around 2.9 eV ± 0.1 eV), or that the magnitude of these shifts fell within the baseline experimental uncertainty of the measurements.

The interaction of water with photocatalytic surfaces governs the water-splitting mechanism. Within this interaction, the direction of surface charge transfer, i.e. whether water acts as a local reducing or oxidizing agent, depends strongly on the material's electronic structure. To probe whether the $Eu^{2+/3+}$ CTL remains electrochemically active under conditions relevant to water splitting, we exposed the calcium-containing oxynitride surfaces to 1 mbar water vapour at room temperature within a Near-Ambient Pressure XPS (NAP-XPS) system (see Figures S8.6 and S8.7). We observe that the introduction of water vapour induces a surface oxidation, driving the conversion of $Eu^{2+}$ to $Eu^{3+}$, as indicated by the attenuation of the $Eu^{2+}$ features in both the Eu 4*d* core levels and the valence band. This behaviour contrasts with observations from Hermans et al. on $BiVO_4$, where room-temperature water exposure raised the Fermi level and reduced surface metal ions ($V^{5+}$ to $V^{4+}$ and $Bi^{3+}$ to $Bi^0$) via reductive water dissociation [42]. This contrasting surface response reflects the different electronic structure of the two materials. In $BiVO_4$, the Fermi level position makes the surface susceptible to reduction by dissociative water adsorption, whereas in the oxynitride, the Fermi level, pinned at the $Eu^{2+/3+}$ charge transition level, seems to favour $Eu^{2+}$ oxidation instead. This observation indicates that similar to influencing nitrogen incorporation in the bulk, the $Eu^{2+/3+}$ charge transition level at the surface also seems to govern surface charge exchange within a water vapour environment.

## 3. Fermi Level Engineering as a tool for oxynitride design

Combining synchrotron XRD, neutron diffraction, 3D electron diffraction, and synchrotron-based XPS reveals that nitrogen incorporation in perovskite-type oxides is regulated by an interplay of the Fermi level position and crystal symmetry. DFT calculations on pristine $BaTiO_3$ show that the incorporation of nitrogen is only energetically favourable at high $E_F$. Introducing europium adds an active redox centre through the $Eu^{2+/3+}$ charge transition level, which lies high in the bandgap [27] and pins the Fermi level to accommodate higher nitrogen incorporation. Nevertheless, the rigid, high-symmetry tetragonal structure of $Eu_{0.2}Ba_{0.8}Ti(O,N)_3$ results in a lower saturation limit, stopping the reaction before the $Eu^{2+}$ reservoir can be fully utilised. The structural rigidity dictated by the large ionic radius of $Ba^{2+}$ ($r_i$ = 1.61 Å) [43] prevents lattice adaptation to volume changes during oxidation of $Eu^{2+}$ to $Eu^{3+}$ ($r_i$ = 1.46 Å to 1.28 Å) [44] and $N^{3-}$ incorporation ($r_i$ = 1.47 Å) [45]. Even though $Eu_{0.8}Ba_{0.2}Ti(O,N)_3$ relaxes into an orthorhombic symmetry, the size mismatch between $Ba^{2+}$ and $Eu^{3+}$ induces massive $Ti(O,N)_6$ octahedra internal distortion, which limits the nitrogen content at roughly N = 0.44 per unit formula.

The orthorhombic Ca-containing samples operate within a more flexible and compliant framework due to the smaller ionic radius of $Ca^{2+}$ ($r_i$ = 1.34 Å) [43], which aligns with the ionic radius of the newly formed $Eu^{3+}$. This decreases atomic strain and enables the $Ti(O,N)_6$ octahedra to tilt and rotate to accommodate the larger $N^{3-}$ anions. In $Eu_{0.2}Ca_{0.8}TiO_3$, the pre-existing distorted orthorhombic oxide minimises the substitution energy barrier upon ammonolysis, enabling stoichiometric nitrogen incorporation (N = 0.2). For $Eu_{0.8}Ca_{0.2}TiO_3$, ammonolysis drives a transition to an orthorhombic oxynitride that restructures the metal-anion orbital overlap and the electronic density of states. The flexible octahedral tilting allows the system to unpin the $E_F$ from the charge transition level, leading to high nitrogen incorporation and a complete $Eu^{2+}$ to $Eu^{3+}$ oxidation without phase segregation.

Our study deliberately initially focuses on pristine $BaTiO_3$-based systems to explore a direct, potentially less environmentally hazardous ammonolysis route. However, the unified concept of Fermi Level Engineering provides a broader viewpoint from which other successful synthetic pathways can be understood. In oxyhydride intermediate precursors ($BaTiO_{3-x}H_x$) [16], the reducing synthesis conditions raise the $E_F$ until titanium reveals mixed $Ti^{4+}/Ti^{3+}$ valence [27], pinning the $E_F$ at this charge transition level to aid controlled nitrogen substitution. Similarly, partial $Ta^{5+}$ substitution on the *B*-site in BTO [18] shifts the $E_F$ toward the CBM prior to nitridation, potentially driving the Fermi level towards a range of more favourable formation energies of nitrogen on oxygen substitution. Viewing distinct experimental approaches through an FLE perspective highlights its potential as a transferable, predictive design tool for effective nitrogen incorporation in perovskites, complementing the classical crystal-chemical guidelines.

## Conclusion

A self-limiting nitridation process in pristine $BaTiO_3$ is revealed, driven by electronic limitations, aligning closely with the concept of Fermi Level Engineering in oxide (electro)ceramics. Increasing nitrogen incorporation shifts the $E_F$ downward, taking it well below the crossing point of acceptor and donor defect formation energies, and making further nitrogen incorporation energetically unfavourable. Consequently, our findings reveal that higher incorporation of nitrogen can be successfully achieved by keeping the $E_F$ at high levels.

We demonstrated that *A*-site cationic co-substitution with Eu and Ca successfully alters the governing defect chemistry. The $Eu^{2+/3+}$ redox couple acts as an electronic buffer that pins the Fermi level at its charge transition level, which lies closer to the CBM. This stabilisation effectively evades the self-limiting process, enabling a more controlled introduction of nitrogen and significantly higher nitrogen concentrations. Therefore, cationic co-substitution shifts the limiting constraints of nitridation from energetic to structural ones, which depend on the symmetry of the crystal structure.

In $Eu_{1-x}Ba_xTi(O,N)_3$, the crystal structure is either rigid, as the tetragonal *P*4*mm* symmetry at high Ba content cannot easily accommodate strain, or the size mismatch between $Ba^{2+}$ and $Eu^{3+}$ induces octahedral distortion in the $Ti(O,N)_6$ polyhedra. As a result, the introduction of the larger $N^{3-}$ ion is structurally constrained, limiting the nitrogen content to N ≈ 0.14 ($x$ = 0.8) and N ≈ 0.44 ($x$ = 0.2), instead of the expected stoichiometric values. Contrarily, $Eu_{1-x}Ca_xTi(O,N)_3$ benefits from a more flexible, orthorhombic structure. The structural tolerance allows for higher lattice distortions minimising that of the $Ti(O,N)_6$ octahedra to accommodate the larger $N^{3-}$ anions, leading to higher nitrogen contents of N ≈ 0.21 ($x$ = 0.8) and N ≈ 0.65 ($x$ = 0.2), and showing the possibility to reach near-stoichiometric oxynitride phases.

Ultimately, the combination of structural and electronic structure characterisation established a clear correlation between nitrogen incorporation, Fermi level positioning, and crystal symmetry. This work's insights provide a predictive framework for future improvements in designing functional oxynitrides via coupled electronic and structural engineering.

## Acknowledgements

We kindly acknowledge the financial support provided by the German Research Foundation (DFG) via the CRC 1548 – FLAIR (Projektnummer: 463184206). The authors deeply appreciate the granted beam time by the Institut Laue-Langevin (ILL), Grenoble, France; and the European Synchrotron Radiation Facility (ESRF), Grenoble, France, and the financial support received during the experiments. Our sincere thanks to Dr. Thomas Hansen (ILL), and Dr. Catherine Dejoie (ESRF) for their kind support during the sample preparation and the neutron diffraction and high-resolution XRD experiments. We are very grateful to the Helmholtz-Zentrum Berlin für Materialien und Energie (HZB, BESSY II) for the allocation of synchrotron radiation beamtime and the technical support of Dr. Ronny Golnak and Dr. Jie Xiao at SoLiAS@OPUS. JPH and PZ acknowledge financial support from the Federal Ministry of Research, Technology and Space (BMFTR), Germany in the framework of ErUM-Pro 05K25RD3. The authors gratefully acknowledge the computing time provided to them at the NHR Center NHR4CES at TU Darmstadt (project number p0023194). This is funded by the Federal Ministry of Research, Technology, and Space, and the state governments participating on the basis of the resolutions of the GWK for national high-performance computing at universities (www.nhr-verein.de/unsere-partner). We would like to thank Dr. Jochen Rohrer, Dr. Lorenzo Villa, and Dr. Sabrina Sicolo (TU Darmstadt) for helpful discussions and advice. Our deep thanks to Dr. Harol Moreno Fernández (Technical University of Darmstadt) for the technical support, assistance with NAP-XPS measurements, and valuable discussions. The authors thank LNNano - Laboratório Nacional de Nanotecnologia for access to the TEM equipment (project no. 20254354; Characterization of Challenging Structures by Total Scattering and Pair Distribution Function (PDF)). We acknowledge the support of João Batista Souza Junior (LNNano, Brazilian Center for Research in Energy and Materials, CNPEM) for the assistance with the TEM measurements and data analysis. We thank Vishnu Mogili for support with the 3DED measurements. NAP-XPS measurements were funded by the Deutsche Forschungsgemeinschaft (DFG, German Research Foundation) under Germany's Excellence Strategy – EXC 2089/2 – 390776260. VS acknowledges financial support from the German Federal Ministry of Research, Technology and Space (BMFTR, Germany), project SINATRA:CO2UPLED/033RC034.

## Supporting Information

### Experimental

Commercial $BaTiO_3$ (Alfa Aesar, 99%) was annealed at $T$ = 1423 K for 12 h. For the oxynitride synthesis, the annealed $BaTiO_3$ was then introduced in a tube furnace to undergo thermal ammonolysis at ambient pressure under flowing ammonia (Westfalen, ≥99.98%) at $T$ = 973 K and $T$ = 1073 K for 4 h using a gas flow rate of 100 mL · min$^{-1}$.

The parent oxides, with general formula $Eu_{1-x}Ba_xTiO_3$ and $Eu_{1-x}Ca_xTiO_3$ ($x$ = 0.2, 0.8), were synthesised via a soft chemistry route and subsequent annealing in forming gas. Stoichiometric amounts of europium(III) oxide (Alfa Aesar, 99.9%), titanium(IV)bis(ammonium dilactato)dihydroxide (Sigma Aldrich, 50 wt.% solution in water), barium acetate (Sigma Aldrich, ≥99%) and calcium acetate monohydrate (Thermo Scientific, 99%) were used to produce the samples. Citric acid (ROTH, ≥99.5%) and ethylene glycol (Merck, ≥99.5%) were used as complexing and gelling agent, respectively. More detailed information regarding synthesis can be found in Xiao et al. [1,2]. After calcination at $T$ = 723 K for 6 h, the white powders were treated in flowing forming gas (AirLiquide, 95 vol.% Ar, 5 vol.% $H_2$; >99.999%) at a flow rate of 300 mL · min$^{-1}$, at $T$ = 1473 K for 12 h. The oxynitrides were synthesised via thermal ammonolysis of the black parent oxides, using a tube furnace under flowing ammonia (Westfalen, ≥99.98%) at ambient pressure and $T$ = 1073 K for 10 h using a gas flow rate of 100 mL · min$^{-1}$.

A STOE STADI MP diffractometer was used to perform the powder X-ray diffraction (XRD) at room temperature in transmission geometry. The instrument utilises a Ge (111) monochromator and Mo-$K_{\alpha1}$ radiation.

Neutron diffraction (ND) experiments were performed using the D20 high-resolution powder diffractometer [3,4] of the Institut Laue-Langevin (ILL) in Grenoble, France, utilising an incident wavelength of 1.120148(1) Å at a take-off angle of 90°, determined by Rietveld refinements of the $Na_2Ca_3Al_2F_{14}$ (NAC) standard sample. The wavelength was chosen to balance neutron flux at the sample and the wavelength-dependent neutron absorption behaviour of Eu [5]. The samples [6] were measured in 5 mm diameter vanadium cans at room temperature.

High-resolution XRD data [7] were collected at the high-flux high-resolution powder X-ray diffractometer using the ID22 beamline of the European Synchrotron Radiation Facility (ESRF) in Grenoble, France, with a wavelength of 0.354432(9) Å [8].

Most of the XRD and ND patterns obtained were refined by the Rietveld method [9,10], using the Fullprof.2k software (version 7.40) [11], taking into account XRD refinements done in previous work [12]. For the refinements, pseudo-Voigt profile functions were used with the Caglioti formula for the full width at half maximum (FWHM) determination. One specific oxide sample's synchrotron XRD data refinements ($Eu_{0.8}Ba_{0.2}TiO_3$) were performed using the Rietveld method with the TOPAS6.0 program [13,14] over the full measured 2$\theta$ range. Instrumental parameters were determined from a reference standard measurement (Si/$LaB_6$) and incorporated into the refinements. The microstructural parameters (crystallite size and strain broadening) were refined to account for the observed peak shapes. Isotropic atomic displacement parameters were constrained where appropriate to minimise parameter correlations and improve refinement stability.

Oxygen and nitrogen contents of all samples were determined via hot gas extraction (HGE) using an Eltra Verder ELEMENTRAC ONH-p2 analyser operated in the ON mode. About 5 mg of each sample was weighed into a nickel capsule and introduced into a nickel basket, used as flux. The capsule was dropped into a graphite crucible containing 50 mg of graphite, promoting complete fusion. The impulse furnace power was 5500 W during the outgassing phase. High-purity helium (99.996 %) was used as carrier gas. Before every measurement, the furnace atmosphere was thoroughly purged with helium to eliminate residual gases and moisture. Calibration was done using certified reference metal samples with well-defined oxygen and nitrogen contents, as follows. Eltra GmbH: Steel 921G, 91100-1003, 0.0036 ± 0.0002 %O, 0.0228 ± 0.0009 %N; Steel 1121R, 91100-1002, 0.0034 %O, 0.0069 %N; Steel 1221S, 91100-1005, 0.0031 ± 0.0003 %O, 0.0537 ± 0.0008 %N; Steel 1022P, 91100-1010, 0.0033 ± 0.0003 %O, 0.155 ± 0.002 %N; Copper 322F, 91000-1003, 0.0251 ± 0.0003 %O; Sigma-Aldrich: $ZrO_2$, 99%, 25.97 %O; Thermo Scientific: $Si_3N_4$, 99.9%, 39.94 %N. Each sample was measured at least three times to ensure reproducibility.

Ultra-high vacuum (UHV) X-ray photoelectron spectroscopy (XPS) measurements were performed inside the DAISY-FUN (Darmstadt Integrated Systems for FUNdamental research) cluster tool, which has a measurement base pressure of approximately $1 \times 10^{-9}$ mbar. The measurements used a monochromated Al $K_\alpha$ X-ray source (XR 50M by SPECS, with $h\nu$ = 1486.74 eV, operating at 12.5 kV), coupled with a Focus 500 monochromator and a PHOIBOS 150 analyser by SPECS. Survey spectra were measured with 20 eV pass energy (step size 0.5 eV) and detail spectra with 10 eV (step size 0.05 eV). The spot size was around 2 mm x 2 mm. The spectrometer's binding energy was calibrated using a sputter-cleaned silver film. Inside the cluster tool, it was possible to heat the samples without air contact. For heating of the samples in $O_2$ partial pressure and ultra-high vacuum, a laser heating system from PREVAC (laser wavelength 976 nm) was used, where the samples were heated from the backside, and a thermocouple contact was fixed at the top of the sample. The laser power was controlled with a PID controller, and liquid nitrogen cooling was used to achieve a stable heating rate (1 K/s). The chosen temperature was then held for 10 minutes, and afterward, the sample cooled down to room temperature.

The near-ambient pressure (NAP)-XPS measurements were performed at the Technical University of Munich. The XPS measurements were performed using a lab-based near-ambient pressure X-ray photoelectron spectroscopy setup (NAP-XPS, Specs Surface Nano Analysis GmbH) in a chamber with a base pressure below $5 \times 10^{-9}$ mbar. A monochromatic Al $K_\alpha$ X-ray source (50 W, 120 µm spot size) provided the excitation. The analyser was calibrated using sputter-cleaned copper, silver, and gold foils, positioning the Au $4f_{7/2}$, Ag $3d_{5/2}$, and Cu $2p_{3/2}$ core-level peaks at 84.0 eV, 368.3 eV, and 932.6 eV, respectively. All spectra were acquired in fixed analyser transmission (FAT) mode with a 0.05 eV step size, 300 ms dwell time, and 30 eV pass energy, achieving a resolution of 0.5 eV. High-resolution spectra for each core level were collected over a minimum of five scans in cyclic mode, each peak scan interleaved with sequential scans of the others. This approach minimised the effect of temporal X-ray power fluctuations on our quantitative analysis. For near-ambient measurements, water vapour was dosed via a mass flow controller (MFC) at a maximum flow rate of 10 ml/min, with chamber pressure held at 1 mbar and regulated by the MFC. All measurements were performed at room temperature.

Synchrotron based XPS and NEXAFS measurements were performed at SoLiAS@OPUS (UE51_PGM-1) at the BESSY II electron storage ring, operated by Helmholtz-Zentrum Berlin für Materialien und Energie. The measurements took place at the SoLiAS endstation, which has a base pressure of approximately $1.5 \times 10^{-10}$ mbar [15]. The system is equipped with a PHOIBOS 150 analyser by SPECS, featuring a 1D DLD detector. The spectrometer's binding energy was calibrated using a sputter-cleaned gold film. Survey spectra were measured with 20 eV pass energy (step size 0.5 eV) and detail spectra with 10 eV (step size 0.05 eV). The spot size was around 100 µm x 150 µm. The incident photon energy was systematically adjusted for all core levels to maintain a consistent kinetic energy of approximately 650 eV for the emitted photoelectrons. This energy was chosen, since it ensures a steady inelastic mean free path (IMFP) throughout all measurements, providing a uniform, surface-sensitive probing depth. Additionally, selecting a kinetic energy of 650 eV effectively reduces spectral overlap with Auger lines and prevents contributions from higher-order harmonics generated by the beamline monochromator. The IMFP for this kinetic energy, calculated for $BaTiO_3$, with the Tanuma-Powell-Penn formula (TPP-2M) [16], is ~1.4 nm, providing an information depth ($d_{95\%}$) of ~4.2 nm, which is expected to be similar for the other compounds. The oxides samples exhibited sufficient intrinsic conductivity to prevent detectable charging effects. In contrast, surface charging was observed during the measurement of the semiconducting oxynitride samples. Therefore, the incident photon flux was intentionally decreased by closing the apertures of the beamline to remove charging artefacts. Furthermore, all core levels exhibited uniform binding energy shifts (see Table S8.1) despite being measured at different photon energies resulting in different incident photon fluxes indicating a successful removal of surface charging. Despite these optimization efforts, there is still a small residual uncertainty in the absolute binding energies of the oxynitrides. XAS spectra were recorded in Auger electron yield (AEY) mode at kinetic energies of 240 eV for the Ti $L_{2,3}$-edge.

All XPS data were analysed using CasaXPS version 2.3.25. The spectra were fitted with a combination of Gaussian-Lorentzian curve (GL(30)), and Shirley background was applied [17].

The SEM-EDX mapping for Eu, Ca, and Ti was performed on a Tescan Mira 3 SEM and an METEK EDAX detector, using an accelerating voltage of 15 kV and beam intensity of 15.

Continuous-rotation three-dimensional electron diffraction (3DED) [18,19] data were collected on a Thermo Scientific Talos Transmission Electron Microscope (TEM) equipped with a field emission gun (FEG) and operated at an accelerating voltage of 200 kV (wavelength $\lambda$ = 0.025082 Å). Powder samples of $Eu_{0.2}Ca_{0.8}Ti(O,N)_3$ were gently crushed in an agate mortar and dispersed in ethanol p.a., and a droplet of the resulting suspension was deposited onto a lacey carbon-coated copper grid. Diffraction patterns were recorded during continuous rotation using a Thermo Scientific Ceta 16M CMOS camera (4096 × 4096 pixels, used here with 2×2 binning, giving an effective frame size of 2048 × 2048 pixels and a pixel size of 0.028 mm) at a spot size of 6, with data acquisition controlled using the EPU-D (Thermo Scientific) software. Frames were acquired over a total tilt range of 120° (from -60° to +60°) at a tilt speed of 2°/s, with a tilt increment of 1° per frame and an exposure time of 0.5 s per frame, at a camera length of 656.4 mm.

Data reduction was performed with PETS2 [20] (calibration constant 0.001701 $Å^{-1}$/pixel), with reflections integrated to a resolution of 0.71 Å ($d$ = 1.4 $Å^{-1}$) using kinematical and dynamical approximations. The structure was solved by direct/dual-space methods using SHELXT [21] and SIR2014 [22], as implemented in Jana2020 [23], and refined using the dynamical refinement procedure in Jana2020.

EDS analyses were carried out using an Oxford SDD X-Max 80 $mm^2$ windowless detector, and the results were verified by EELS measurements performed with a Gatan GIF Tridiem 863 on a JEOL JEM-2100F.

Ultra-violet−visible diffuse reflectance spectroscopy (UV−Vis DRS) data were recorded at room temperature over a wavelength range of 240−2000 nm (5.2−0.6 eV) using a J&M TIDAS UV−Vis−NIR spectrophotometer equipped with a reflection measuring head. A $BaSO_4$ standard was used as a reference. Further details regarding the method used for the estimation of the optical bandgap can be found in section S3.

**Computational details**

Density functional theory (DFT) calculations were performed using the Vienna Ab Initio Simulation Package (VASP) [24–26]. The projector-augmented wave (PAW) scheme [27,28] described the electron-ion interaction. For the valence electron configuration, Ba was considered with $5s^2 5p^6 5d^{0.01} 6s^{1.99}$, Ti with $3s^2 3p^6 3d^3 4s^1$, O with $2s^2 2p^4$.

Calculations with hybrid functionals were performed to obtain an accurate reference band gap for the 0 K-stable trigonal rhombohedral phase of $BaTiO_3$ ($R3m$, space group No. 160). The band structure was calculated using the Heyd-Scuseria-Ernzerhof (HSE06) functional [29,30], with a mixing of 16% and a screening parameter of 0.2 $Å^{-1}$, employing a 5-atom unit cell. The wave functions were expanded into plane waves up to an energy cutoff of 750 eV. The Brillouin zone was sampled using a $\Gamma$-centered $6 \times 6 \times 6$ $k$ mesh. The atomic positions were relaxed until the Hellmann-Feynman forces acting on each atom were below 0.05 eV/Å.

For the defect calculations, the electron exchange and correlation effects were treated within the generalised gradient approximation (GGA) using the Perdew-Burke-Ernzerhof (PBE) [31] functional, using a $3 \times 3 \times 3$ supercell containing 135 atoms. The same plane-wave cutoff and relaxation criterion were used as for the HSE06 reference calculations; however, the Brillouin zone was sampled using a $\Gamma$-centered $2 \times 2 \times 2$ $k$ mesh for the supercell.

The lattice parameters and energy gaps calculated using both functionals are reported in Table S 1.

Table S 1. Values of lattice constant *a* (Å), angle α (°), and bang gap $E_g$ of rhombohedral $BaTiO_3$ (R3m, space group No. 160), computed with the PBE [27] and HSE06 [32,33] exchange-correlation functionals and confronted with experimental data from Kwei et al. [34]. A measured band gap for this phase was not established from the references examined.

| | ***a* (Å)** | **α (°)** | **$E_g$ (eV)** |
|---:|:---:|:---:|:---:|
| PBE | 4.0753 | 89.748 | 2.29 |
| HSE06 | 4.0440 | 89.702 | 3.55 |
| Exp. [34] | 4.00360 | 89.839 | – |

The defect thermodynamics applied in this work can be found down below, in section S1.

# S1. Density Functional Theory (DFT)

## S1.1. Defect Thermodynamics

The formation energy $\Delta E_D^f$ of a defect $D$ with charge $q$ can be written as [35]:

$$\Delta E_D^f = E_D - E_P + q(E_{VBM} + \mu_e) + E_{corr} - \sum_i \Delta n_i \mu_i, \tag{S1}$$

where $E_D$ is the total energy of the system with defect $D$ and charge $q$, and $E_P$ is the energy of the pristine reference cell. The third term describes the dependence of the formation energy on the electron chemical potential $\mu_e$, while the position of the valence band maximum (VBM) $E_{\text{VBM}}$ is taken as its reference. $E_{\text{corr}}$ describes the applied energy corrections. The spurious electrostatic interactions between periodic images of charged defects, which arise due to the finite size of the supercell, are corrected using the scheme proposed by Kumagai *et al.* [32]. It is well known that semi-local functionals, such as PBE, severely underestimate the band gap and can, in some cases, introduce uncertainties in calculated defect energetics [35]. To account for this, the band gap is corrected by rigidly shifting the valence band maximum (VBM) and the conduction band minimum (CBM) as obtained from PBE calculations to the respective values obtained by HSE06 [29,30] calculations. The correction energy is then given by:

$$\Delta E_{bg} = n_e \Delta E_{\mathrm{CBM}} + n_h \Delta E_{\mathrm{VBM}}, \quad \text{(S2)}$$

where $n_e$ and $n_h$ are the number of electrons introduced in the conduction band and the number of holes introduced in the valence band, respectively. $\Delta E_{\mathrm{CBM}}$ and $\Delta E_{\mathrm{VBM}}$ are the shifts in the CBM and VBM between HSE06 and PBE. Here, donor defects follow the energy correction of the CBM, while acceptor defects follow the correction of the VBM. Since the VBM is shifted downward, $E_{\mathrm{VBM}}$ is also shifted by:

$$E_{\mathrm{VBM}} = E_{\mathrm{VBM}}^{\mathrm{PBE}} - \Delta E_{\mathrm{VBM}}. \quad \text{(S3)}$$

Finally, the last term accounts for the dependence of the formation energy on the chemical potentials of the constituent elements. Here, $\Delta n_i$ is the difference between the number of atoms of species $i$ in the defective and bulk systems. The chemical potential $\mu_i$ of element $i$ can be rewritten as:

$$\mu_i = \mu_i^{\mathrm{ref}} + \Delta\mu_i, \quad \text{(S4)}$$

where $\mu_i^{\mathrm{ref}}$ is the chemical potential of the element $i$ in its stable elemental phase.

The Density Functional Theory data used to generate the phase stability diagram for rhombohedral $BaTiO_3$ were obtained from the Materials Project (MP) database (PYMATGEN implementation, version 2025.5.2) [33,36,37], which uses consistent computational parameters with respect to our defect calculations. It is well known that semi-local DFT (e.g., LDA or GGA) exhibits an overbinding error in the $O_2$ and $N_2$ molecule [38–40]. To account for this effect, all entries used to construct the phase diagram were processed using the Materials Project compatibility corrections as implemented in PYMATGEN (class MaterialsProjectCompatibility). These corrections do not address the $O_2$ molecule directly. Instead, they apply a correction to oxygen-containing compounds (with distinct values for oxide, peroxides, and superoxides, depending on the oxygen bonding environment) [41].

A charge transition level (CTL), $\varepsilon$, is the electron chemical potential at which the formation energy of a given defect in two different charge states ($q_1$and $q_2$) becomes equal. It is given by:

$$\varepsilon = -\frac{\Delta E_D^F(q_1, \mu_e = 0) - \Delta E_D^F(q_2, \mu_e = 0)}{q_1 - q_2}, \quad \text{(S5)}$$

where $\Delta E_D^F(q_1, \mu_e = 0)$ and $\Delta E_D^F(q_2, \mu_e = 0)$ are the defect formation energies at charge states $q_1$ and $q_2$, respectively, when the electron chemical potential $\mu_e$ is at the VBM.

At thermodynamic equilibrium, the concentration of a defect $D$ in charge state $q$ is given by:

$$[D_q] = N_{\text{sites}} \exp\left(-\frac{\Delta E^F_{q,D}(\mu_e)}{k_B T}\right), \tag{S6}$$

where $q$ represents the defect charge, $N_{\text{sites}}$ the concentration of the corresponding lattice sites, $T$ the absolute temperature, and $k_B$ the Boltzmann constant.

It is important to stress that the electron chemical potential $\mu_e$ is not a free parameter. It is rather determined by the charge neutrality condition:

$$n_e + n_A = n_h + n_D, \tag{S7}$$

in which $n_e$ and $n_h$ are the electron and hole carrier concentrations, respectively, while $n_D$ and $n_A$ are the donor and acceptor concentrations, respectively.

Considering all defect species $D$ in their different charge states $q$, Equation S7 can be reformulated as:

$$\sum_{q,D} q \cdot [D_q] - n_e + n_h = 0. \tag{S8}$$

Intrinsic carrier concentrations are computed by integrating the number of unoccupied states up to the VBM ($E_{\text{VBM}}$) for holes,

$$n_h = \int_{-\infty}^{E_{\text{VBM}}} D(\varepsilon)[1 - f(\varepsilon)]\mathrm{d}\varepsilon, \tag{S9}$$

and the number of occupied states from the conduction band minimum (CBM) ($E_{\text{CBM}}$) for electrons,

$$n_e = \int_{E_{\text{CBM}}}^{+\infty} D(\varepsilon) f(\varepsilon)\mathrm{d}\varepsilon, \tag{S10}$$

where $f(\varepsilon, \mu_e) = [\exp(\varepsilon - \mu_e)/k_B T]^{-1}$ is the Fermi-Dirac distribution.

In this study, the integrals in Equation S9 and S10 were computed as Riemann sums over the energies in the calculated density of states, following the method delineated by PYMATGEN [36,37]. Subsequently, the equilibrium electron chemical potential $\mu_e$ was determined self-consistently by solving the charge neutrality condition for $\mu_e$ at a given temperature.

### S1.2. Oxygen and nitrogen chemical potentials

Defect formation energies depend on the chemical potential of the reservoir. For oxygen, $\mu_{\text{O}}$ varies with its partial pressure $p_{\text{O}_2}$ and can be expressed via the ideal gas law as:

$$\mu_{\text{O}}\left(T, p_{\text{O}_2}\right) = \mu_{\text{O}}(T, p^0) + \frac{1}{2} k_B T \ln\left(\frac{p_{\text{O}_2}}{p^0}\right). \tag{S11}$$

Here, $k_B$ is the Boltzmann constant, $T$ the temperature, $\mu_{\text{O}}$ the oxygen chemical potential, and $p^0$ the reference pressure.

The computed energy of the $O_2$ molecule at 0 K $\mu_{\text{O}}(0\ \text{K}, p^0)$ = −4.95 eV serves as the reference. The temperature dependence $\mu_{\text{O}}(T, p^0)$ was calculated following Reuter and Scheffler's methodology [42], using the thermochemical tables in [43]. This procedure enables a mapping between the experimentally controlled variables, such as temperature and oxygen partial pressures, and the oxygen chemical potential used in our defect-thermodynamic model. For nitrogen, the 0 K reference was taken from the legacy Materials Project MaterialsProjectCompatibility scheme, where the $\mu_{\mathrm{N}}(0\ \mathrm{K}, p^0) = -7.95\ \mathrm{eV}$ reference is fitted to experimental formation energies of various nitrides. Although nitrogen is experimentally introduced by annealing in an ammonia-containing atmosphere, the high treatment temperatures considered here are assumed to be sufficient for ammonia to decompose and equilibrate with molecular nitrogen and hydrogen [44]. Therefore, the nitrogen chemical potential was referenced to $N_2$, rather than directly to $NH_3$.

Using this mapping, defect concentrations and the electron chemical potential can be evaluated as functions of the gas partial pressures within a consistent theoretical framework, facilitating comparison of defect-chemistry trends with experiment. We note that additional finite-temperature contributions to the free energies are neglected. Therefore, absolute concentrations should be interpreted with appropriate caution at high temperature.

To define the thermodynamic reservoirs, we constructed the semigrand canonical phase diagram, which maps the stability regions as a function of the chemical potentials of the constituents. The list of stable phases is reported in Table S 1.1.

Table S 1.1. Formation energy per atom for all the materials used to compute the stability diagram. The data is taken from the Materials Project database (PYMATGEN implementation, version 2025.5.2) [33,36,37]

| Composition | Structure | $\Delta E^{\mathrm{f}}$ (eV/atom) |
|---|---|---|
| Ba | Cubic ($Im\bar{3}m$) | 0.00 |
| $Ba_2Ti_{13}O_{22}$ | Orthorhombic ($Cmce$) | -3.38 |
| $Ba_2Ti_6O_{13}$ | Monoclinic ($C2/m$) | -3.50 |
| $Ba_2TiO_4$ | Monoclinic ($P2_1/c$) | -3.41 |
| $Ba_3Ti_{20}O_{40}$ | Monoclinic ($C2/m$) | -3.49 |
| $Ba_4Ti_{13}O_{30}$ | Orthorhombic ($Cmce$) | -3.54 |
| BaO | Cubic ($Fm\bar{3}m$) | -2.83 |
| $BaO_2$ | Tetragonal ($I4/mmm$) | -2.17 |
| $BaTiO_5$ | Orthorhombic ($Pnma$) | -3.54 |
| $BaTi_8O_{16}$ | Monoclinic ($C2$) | -3.49 |
| $BaTiO_3$ | Trigonal ($R3m$) | -3.50 |
| $O_2$ | Monoclinic ($C2/m$) | 0.00 |
| Ti | Hexagonal ($P6/mmm$) | 0.00 |
| $Ti_2O$ | Trigonal ($P\bar{3}m1$) | -2.04 |
| $Ti_2O_3$ | Trigonal ($R\bar{3}c$) | -3.31 |
| $Ti_3O$ | Trigonal ($P\bar{3}1c$) | -1.57 |
| $Ti_3O_5$ | Monoclinic ($C2/m$) | -3.39 |
| $Ti_6O$ | Trigonal ($P\bar{3}1c$) | -0.92 |
| TiO | Hexagonal ($P\bar{6}2m$) | -2.92 |
| $TiO_2$ | Monoclinic ($C2/m$) | -3.51 |

The resulting stability diagram, obtained from these formation energies, is shown in Figure S 1.1. The allowed range of chemical potential is constrained by the condition that their sum must be equal to the formation enthalpy of $BaTiO_3$:

$$\Delta\mu_{\mathrm{Ba}} + \Delta\mu_{\mathrm{Ti}} + 3\Delta\mu_{\mathrm{O}} = \Delta H_f[\mathrm{BaTiO_3}] \qquad \text{(S12)}$$

For a fixed sum of $\Delta\mu_{\mathrm{Ba}} + \Delta\mu_{\mathrm{Ti}}$, the oxygen chemical potential $\Delta\mu_{\mathrm{O}}$ is determined by this condition. The stability region of $BaTiO_3$ is further limited by the presence of competing phases and is represented by the area delimited by points A to D, in Figure S 1.1. More specifically, given a generic competing phase of composition $Ba_iTi_jO_k$, the chemical potentials must satisfy the condition:

$$i\Delta\mu_{\mathrm{Ba}} + j\Delta\mu_{\mathrm{Ti}} + k\Delta\mu_{\mathrm{O}} \leq \Delta H_f\left[\mathrm{Ba}_i\mathrm{Ti}_j\mathrm{O}_k\right]. \tag{S13}$$

The defect formation energies were calculated for different thermodynamic reservoirs, represented in Figure S 1.1. by the labels from A to D and X. Regions A-D span from O-rich conditions (points C and D) to Ba-rich and Ti-rich conditions (points A and B).

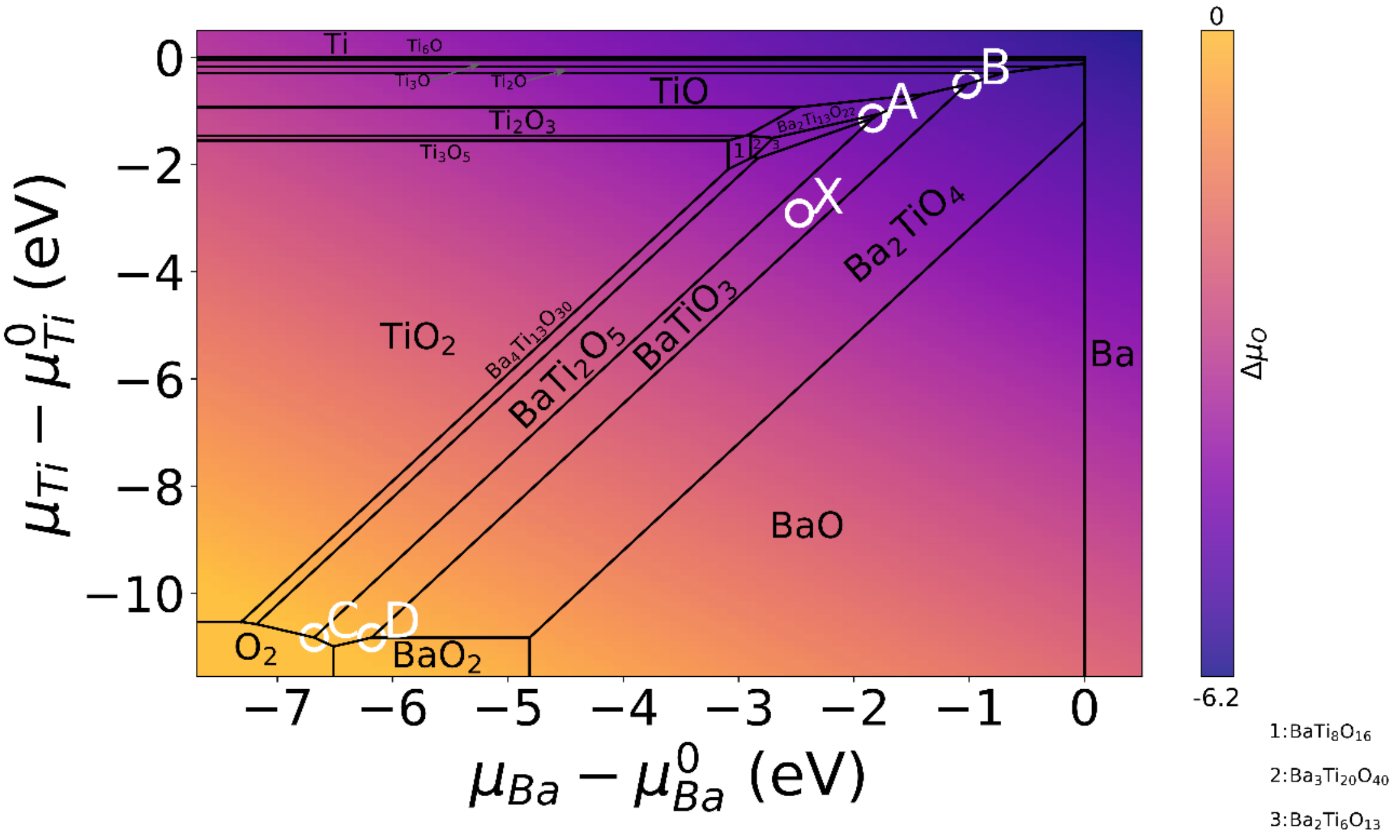

Figure S 1.1. Stability diagram of the ternary Ba-Ti-O system, as derived from the data in Table S 1.1. The defect formation energies are discussed in terms of the chemical potentials at points A to D and point X, as reported in Table S 1.2.

We also evaluated the formation energies using a thermodynamic reservoir that represents the experimental conditions. In particular, we fix the oxygen chemical potential ($\Delta\mu_{\mathrm{O}}$) to its value under the conditions of temperature and partial pressure of the nitridation process ($T = 1073\ \mathrm{K}, p_{\mathrm{O}_2} = 10^{-27}$ atm). At these finite-temperature conditions, the $0\ \mathrm{K}$-stability diagram is kept fixed, and temperature/pressure enter only through Equation S9, which details the calculation of this quantity. With this procedure, we obtained $\Delta\mu_{\mathrm{O}} = -4.04\ \mathrm{eV}$, which is indicated by the point X in Figure S 1.1.

For each oxygen partial pressure, we first determined $\mu_{\mathrm{O}}$. The remaining cation chemical potentials were then constrained by $BaTiO_3$ phase stability in the Ba-Ti-O system of Figure S 1.1: at fixed $\mu_{\mathrm{O}}$, the allowed $\mu_{\mathrm{Ba}}$ and $\mu_{\mathrm{Ti}}$ values form a line segment given by the intersection of the constant $\mu_{\mathrm{O}}$ line with the $BaTiO_3$ stability region. The endpoints of this segment correspond to the limiting equilibria $BaTiO_3$ + $BaTi_2O_5$ (Ti-rich limit) and $BaTiO_3$ + $Ba_2Ti_2O_4$ (Ba-rich limit). We adopted a representative condition defined as the midpoint between these endpoints for each $\mu_{\mathrm{O}}$, and used the resulting set to compute defect formation energies and solve the charge-neutrality condition.

Assuming complete decomposition of $NH_3$ according to $2NH_3 \rightarrow N_2 + 3H_2$, the stoichiometric product gas contains a mole fraction $x_{N_2} = 0.25$. At a total pressure of 1 atm, this corresponds to $p_{N_2} = 0.25$ atm. We therefore use $N_2$ as an effective nitrogen reservoir, with $\Delta\mu_N = -1.18$ eV.

Since the range of stability varies for every compound, we evaluated the chemical potential of the elements in O-poor and O-rich regions, assigning the former to reservoirs A and B and the latter to reservoirs C and D. In region X, the values were taken from the center of the stability region, having fixed $\Delta\mu_O$ to $-4.04$ eV. This procedure was employed to compare the formation energies of intrinsic defects. The chemical potential values associated with the labelled regions in the stability diagram are reported in Table S 1.2.

Table S 1.2. Values in eV of the chemical potentials of the individual elements for each of the labelled regions in the stability diagram Figure S 1.1.

| ***Reservoir*** | $\Delta\mu_{Ba}$ | $\Delta\mu_{Ti}$ | $\Delta\mu_O$ | $\Delta\mu_N$ |
|---|---|---|---|---|
| A | -1.84 | -1.14 | -4.84 | -1.18 |
| B | -1.02 | -0.50 | -5.33 | -1.18 |
| C | -6.68 | -10.83 | 0.00 | -1.18 |
| D | -6.18 | -10.83 | -0.17 | -1.18 |
| X | -2.47 | -2.91 | -4.04 | -1.18 |

To assess the contribution of different point defects to the defect chemistry of $BaTiO_3$ under the investigated conditions, we compared the equilibrium defect concentrations obtained from the full defect model with those calculated considering only $N_O$ and $v_O$ (Figure S 1.2). The full model includes $N_O$, $v_O$, $v_{Ba}$, and $v_{Ti}$. As illustrated by the formation energies in Figure S 1.2.c, the cation vacancies exhibit comparatively high formation energies, resulting in negligible equilibrium concentrations. Consequently, their exclusion from the charge-neutrality calculation is expected to have little influence on the concentrations of the dominant defects.

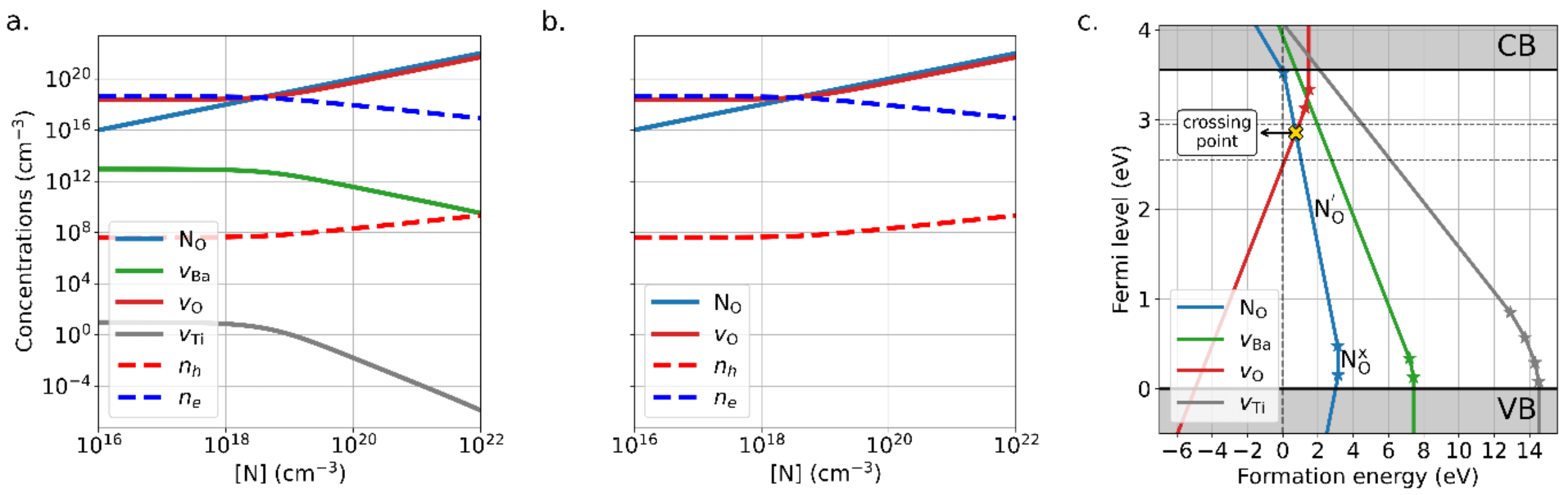

Figure S 1.2. a. Concentrations of all considered defect species as a function of imposed nitrogen concentration [N], obtained from the charge-neutrality condition at $T$ = 1073 K. b. Corresponding defect concentrations recalculated considering only $N_O$ and $v_O$. c. Formation energies of all considered point defects as a function of $E_F$, plotted with transposed axes. Stars indicate the corresponding charge transition levels, and the marked crossing point indicates equal formation energies of $N_O$ and $v_O$. The high formation energies and low concentrations of $v_{Ba}$ and $v_{Ti}$ under these conditions motivate their exclusion from the reduced defect model discussed in the main text. Formation energies were determined at $T$ = 1073 K and $p_{O_2}$ = $10^{-27}$ atm, with the nitrogen chemical potential corresponding to an ideal-gas $N_2$ reservoir with $p_{N_2}$ = 0.25 atm, estimated assuming complete $NH_3$ dissociation at a total pressure of 1 atm.

## S2. Crystallography Data – BaTi(O,N)$_3$

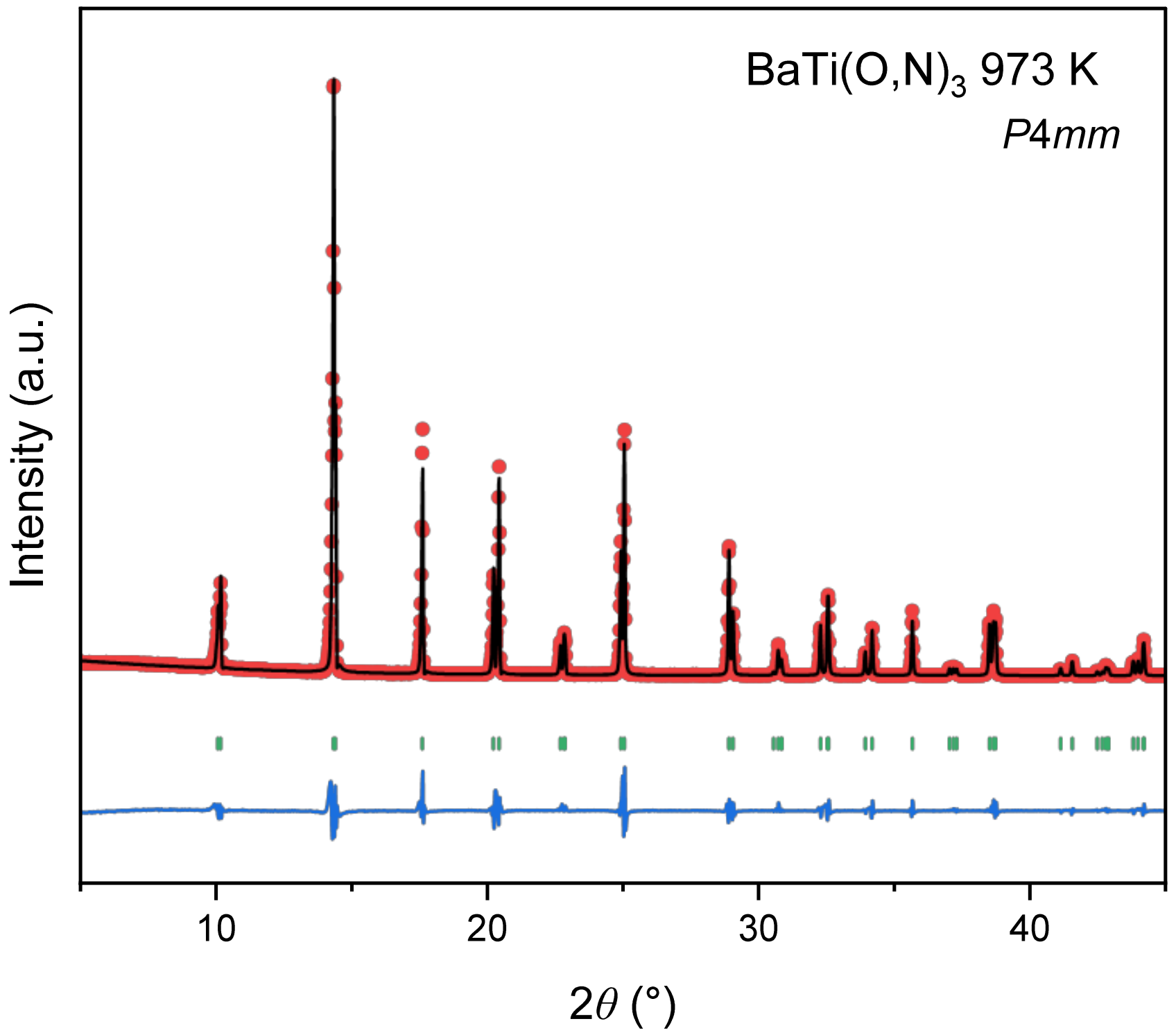

Figure S 2.1. Refined XRD ($\lambda$ = 0.70932 Å) pattern of BaTi(O,N)$_3$ synthesised at $T$ = 973 K, using the tetragonal *P*4*mm* space group model. The red dots represent the experimental data, the black line the calculated pattern, and the blue line the difference curve. The green hash marks represent the indexed reflections of the model.

Table S 2.1. Summary of measurement conditions, unit cell parameters, and goodness of fit values for $BaTiO_3$ and $BaTi(O,N)_3$ synthesised at $T$ = 973 K and $T$ = 1073 K.

| | $BaTiO_3$ | $BaTi(O,N)_3$ 973 K | $BaTi(O,N)_3$ 1073 K |
|---|---|---|---|
| Measurement type | XRD | XRD | XRD |
| Wavelength $\lambda$ (Å) | 0.70932 | 0.70932 | 0.70932 |
| Measurement range | 5° ≤ $2\theta$ ≤ 47° | 5° ≤ $2\theta$ ≤ 47° | 5° ≤ $2\theta$ ≤ 47° |
| Space group | *P*4*mm* | *P*4*mm* | *P*4*mm* |
| $a$ (Å) | 3.99753(2) | 3.99861(2) | 3.99838(3) |
| $b$ (Å) | = $a$ | = $a$ | = $a$ |
| $c$ (Å) | 4.03683(3) | 4.03814(4) | 4.03658(4) |
| $V$ (Å³) | 64.509(1) | 64.565(1) | 64.533(1) |
| Overall $B_{iso}$ (Å$^2$) | 0.32(3) | 0.24(3) | 0.25(3) |
| $R_p$ (%) | 9.92 | 9.87 | 10.2 |
| $R_{wp}$ (%) | 15.1 | 14.7 | 15.2 |
| $R_{Bragg}$ (%) | 4.00 | 4.26 | 4.07 |
| $\chi^2$ | 42.7 | 21.3 | 23.3 |

## S3. Optical Properties – $BaTi(O,N)_3$

The optical response was assessed by calculating apparent absorbance as:

$$A = -\log(R) \tag{S14}$$

where $A$ is absorbance and $R$ is reflectance, and by plotting the Kubelka−Munk (K−M) function $F(R\infty)$ directly against wavelength. The optical bandgaps of the samples were estimated using the K−M function [45], which transforms diffuse reflectance data into absorbance-like values suitable for powdered samples [46]. The reflectance of an optically thick sample ($R\infty$) was converted using the relation:

$$F(R\infty) = (1 - R\infty)2/(2R\infty) \tag{S15}$$

where $F(R\infty)$ approximates the absorbance. Kubelka–Munk-transformed spectra $F(R\infty)$ were plotted as a function of photon energy for the entire sample series without normalisation in order to retain absolute differences in reflectance-derived absorption intensity.

In this study, band gaps were determined by plotting photon energy ($hv$) directly against the Kubelka−Munk absorbance function $F(R\infty)$, i.e., 'energy vs absorbance' plot, and estimating the onset of absorption from the inflection point or linear rise of the absorption edge of the raw, non-normalized spectra. Direct and indirect optical transitions were further evaluated using Tauc plots constructed from Kubelka−Munk-transformed diffuse reflectance data, with normalisation applied only for visual comparison of spectral features. Bandgap onset energies were extracted from the linear absorption edge of the raw Kubelka–Munk spectra, while normalised spectra were used exclusively for comparison of spectral shapes and absorption-edge evolution.

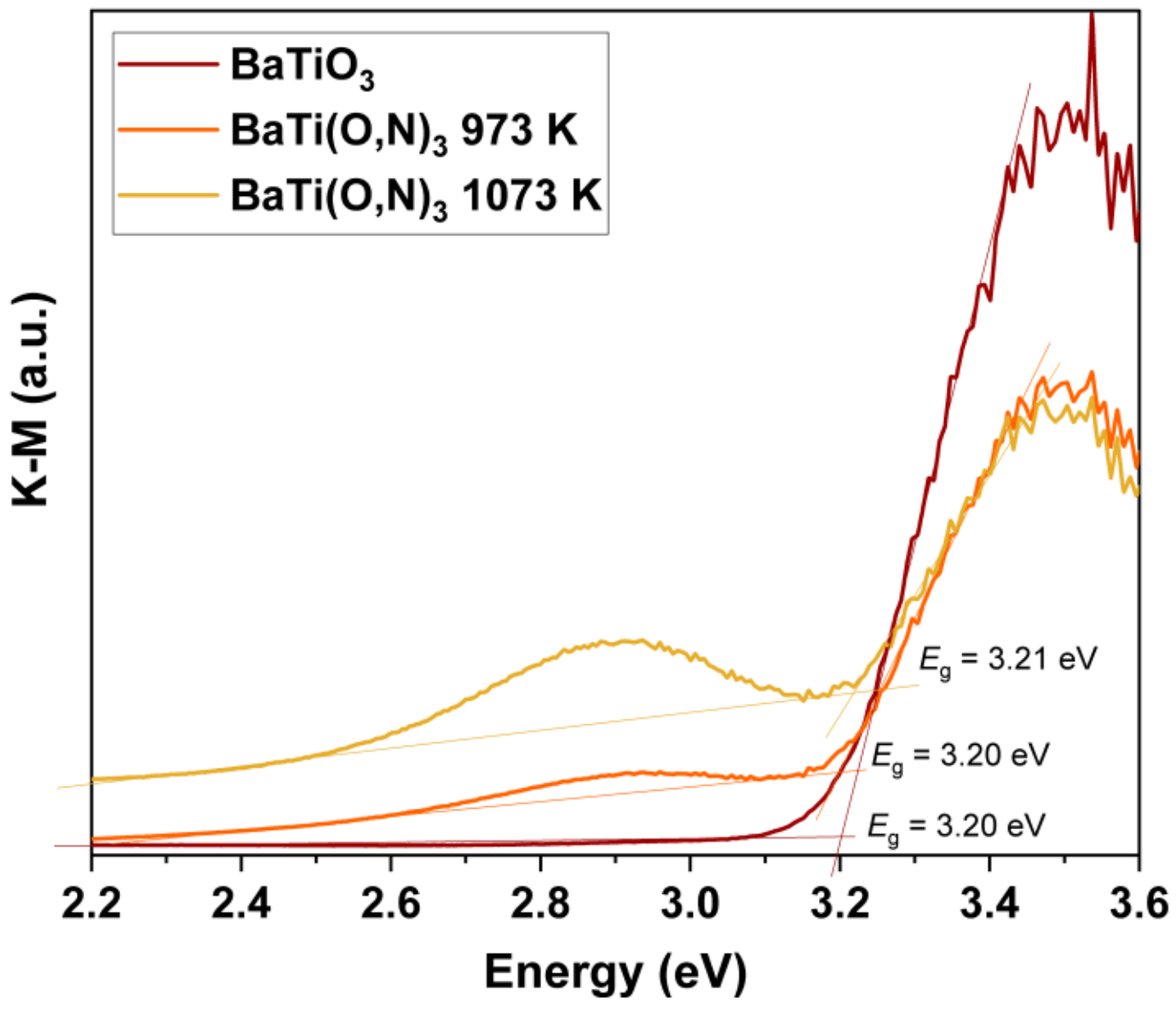

Figure S 3.1. UV-Vis Diffuse Reflectance Spectra, plotted in K-M vs Energy, for $BaTiO_3$ and resulting oxynitrides.

## S4. Electronic Structure Data – $BaTi(O,N)_3$

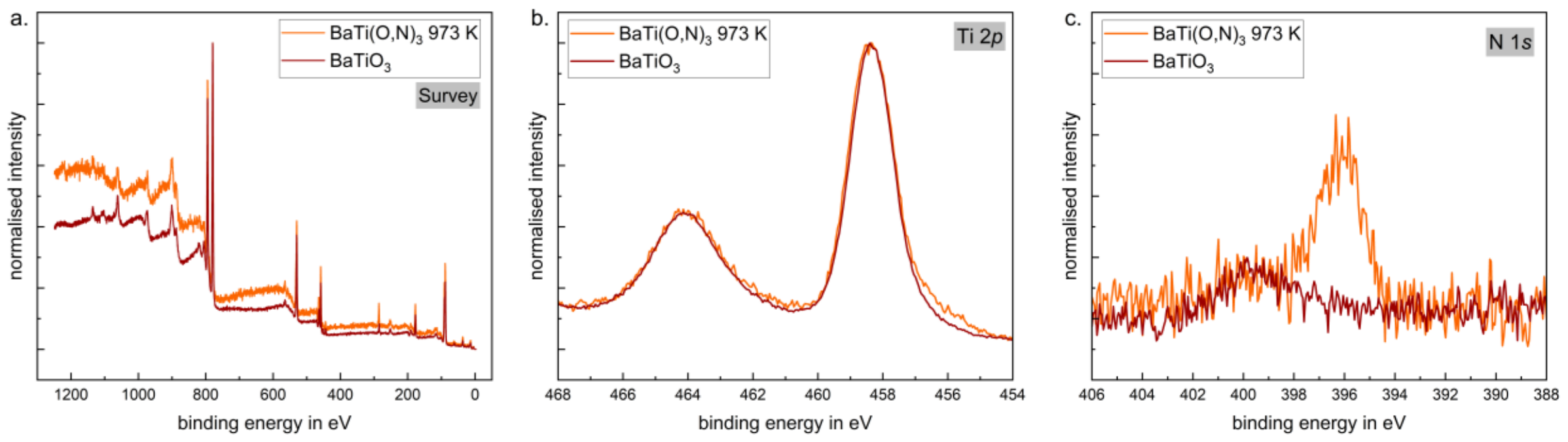

Figure S 4.1 X-ray photoelectron spectra of a. the survey scans and b. corresponding Ti 2*p* and c. N 1*s* core levels of $BaTiO_3$ and BaTi(O,N) measured at DAISY-FUN. Charge compensation with an electron flood gun was applied.

## S5. Crystallography Data – $Eu_{1-x}Ba_xTi(O,N)_3$ / $Eu_{1-x}Ca_xTi(O,N)_3$

### S5.1. Neutron Diffraction Refinements

For all oxynitrides, the thermal atomic displacement parameters were refined as an overall value, and not as independent values. The atomic positions were kept fixed at the values obtained from the refinements of the synchrotron XRD data of the precursor oxides. The Ti occupancies were kept at nominal values, and Eu, Ba, and Ca occupancies were kept at the values refined from the parent oxides. Finally, for $Eu_{0.8}Ba_{0.2}Ti(O,N)_3$, given the very low signal-to-noise ratio of the data due to the neutron absorption of Eu, symmetry lowering, and large reflection broadening, the atomic positions were kept fixed at the nominal values. This was done in order to avoid strong parameter correlation.

Table S 5.1. Structural data from Rietveld refinements of the neutron diffraction patterns of $Eu_{0.2}Ba_{0.8}Ti(O,N)_3$ and $Eu_{0.8}Ba_{0.2}Ti(O,N)_3$. A secondary phase (TiN) is observed for the first. The atomic positions were kept fixed at theoretical and at the values obtained from the refinements of the synchrotron XRD data of the precursor oxide, seen in Table S 5.6. The Ti, and Eu, Ba occupancies were kept at nominal and at the values refined from the parent oxides, respectively. For $Eu_{0.8}Ba_{0.2}Ti(O,N)_3$, given the very low signal-to-noise ratio of the data due to the neutron absorption of Eu, the atomic positions were kept fixed at the nominal values. Occ = occupancy. For the *P*4*mm* structure, (O1)/N(1) occupy the 1*b* Wyckoff position, and O(2)/N(2) occupy the 2*c* position. For the *Pnma* structure, O(1)/N(1) occupy the 4*c* Wyckoff position, and O(2)/N(2) occupy the 8*d* position.

| | $Eu_{0.2}Ba_{0.8}Ti(O,N)_3$ | | $Eu_{0.8}Ba_{0.2}Ti(O,N)_3$ |
|---|---|---|---|
| Phase | oxynitride | TiN | oxynitride |
| Measurement type | PND | | PND |
| Wavelength $\lambda$ (Å) | 1.120148(1) | | 1.120148(1) |
| Measurement range | 0° ≤ 2$\theta$ ≤ 153° | | 0° ≤ 2$\theta$ ≤ 153° |
| Space group | *P*4*mm* | *Fm*$\bar{3}$*m* | *Pnma* |
| Fraction (wt.%) | 98.8(2) | 1.2(2) | 100 |
| *a* (Å) | 3.9785(2) | 4.167(2) | 5.538(3) |
| *b* (Å) | 3.9785(2) | = *a* | 7.842(4) |
| *c* (Å) | 3.9756(5) | = *a* | 5.480(4) |
| *V* (Å³) | 62.927(9) | 72.33(5) | 238.0(3) |
| Overall $B_{iso}$ (Å$^2$) | 0.57(2) | -- | 0.2(1) |
| Occ((O1)/(N1)) | 0.680/0.136(1) | | 0.89/0.11(9) |
| Occ((O2)/(N2)) | 0.924/0.024(1) | | 1.00/0.00(5) |
| $R_p$ (%) | 1.63 | | 6.69 |
| $R_{wp}$ (%) | 2.07 | | 9.65 |
| $R_{Bragg}$ (%) | 19.1 | 42.0 | 1.97 |
| $\chi^2$ | 2.17 | | 1.27 |

Table S 5.2. Structural data from Rietveld refinements of the neutron diffraction patterns of $Eu_{0.2}Ca_{0.8}Ti(O,N)_3$ and $Eu_{0.8}Ca_{0.2}Ti(O,N)_3$. A secondary phase (TiN) is observed for the first. The atomic positions were kept fixed at the values obtained from the refinements of the synchrotron XRD data of the precursor oxides, seen in Table S 5.8 and Table S 5.9. The Ti occupancies were kept at nominal values, and Eu and Ca occupancies were kept at the values refined from the parent oxides, respectively. Occ = occupancy. For both *Pnma* structures, O(1)/N(1) occupy the 4*c* Wyckoff position, and O(2)/N(2) occupy the 8*d* position.

| | $Eu_{0.2}Ca_{0.8}Ti(O,N)_3$ | | $Eu_{0.8}Ca_{0.2}Ti(O,N)_3$ |
|---|---|---|---|
| Phase | oxynitride | TiN | oxynitride |
| Measurement type | PND | | PND |
| Wavelength $\lambda$ (Å) | 1.120148(1) | | 1.120148(1) |
| Measurement range | 0° ≤ $2\theta$ ≤ 153° | | 0° ≤ $2\theta$ ≤ 153° |
| Space group | *Pnma* | $Fm\bar{3}m$ | *Pnma* |
| Fraction (wt.%) | 98(1) | 1.7(1) | 100 |
| $a$ (Å) | 5.4497(1) | 4.2174(4) | 5.5020(3) |
| $b$ (Å) | 7.6654(2) | = $a$ | 7.7585(4) |
| $c$ (Å) | 5.4049(1) | = $a$ | 5.4850(4) |
| $V$ (Å³) | 225.79(1) | 75.01(1) | 234.14(3) |
| Overall $B_{iso}$ (Å$^2$) | 0.56(2) | -- | 0.53(3) |
| Occ((O1)/(N1)) | 0.79/0.14(1) | | 0.682/0.304(1) |
| Occ((O2)/(N2)) | 0.88/0.11(1) | | 0.760/0.142(1) |
| $R_p$ (%) | 3.04 | | 2.57 |
| $R_{wp}$ (%) | 3.98 | | 3.24 |
| $R_{Bragg}$ (%) | 8.91 | 10.1 | 12.8 |
| $\chi^2$ | 11.2 | | 10.4 |

Table S 5.3. Atomic positions data from the Rietveld refinements of the neutron diffraction pattern of $Eu_{0.2}Ca_{0.8}Ti(O,N)_3$.

| | $x$ | $y$ | $z$ |
|---|---|---|---|
| Eu/Ca | 0.5133(8) | $^1/_4$ | -0.001(2) |
| Ti | 0 | 0 | 0 |
| O(1)/N(1) | 0.500(1) | $^1/_4$ | 0.5350(6) |
| O(2)/N(2) | 0.2773(6) | 0.0359(3) | 0.2173(7) |

Table S 5.4. Bond angles and lengths in all three *Pnma* oxynitrides; *X*(1) is located on the Wyckoff site 4*c* and *X*(2) on 8*d*, respectively.

| | $Eu_{0.8}Ba_{0.2}Ti(O,N)_3$ | $Eu_{0.8}Ca_{0.2}Ti(O,N)_3$ | $Eu_{0.2}Ca_{0.8}Ti(O,N)_3$ |
|---|---|---|---|
| ∢(*B*–*X*(1)–*B*) (°) | 158.76(1) | 168.70(1) | 158.23(1) |
| ∢(*B*–*X*(2)–*B*) (°) | 169.7(9) | 158.8(2) | 158.82(1) |
| *d*(*B*–*X*(1)) (Å) | 1.995(3) | 1.9491(3) | 1.9515(3) |
| *d*(*B*–*X*(2)) (Å) | 1.92(1) | 1.996(4) | 1.949(1) |
| Δ*d* (%) | 3.76 | 2.35 | 0.13 |

## S5.2. High Resolution Synchrotron XRD Refinements

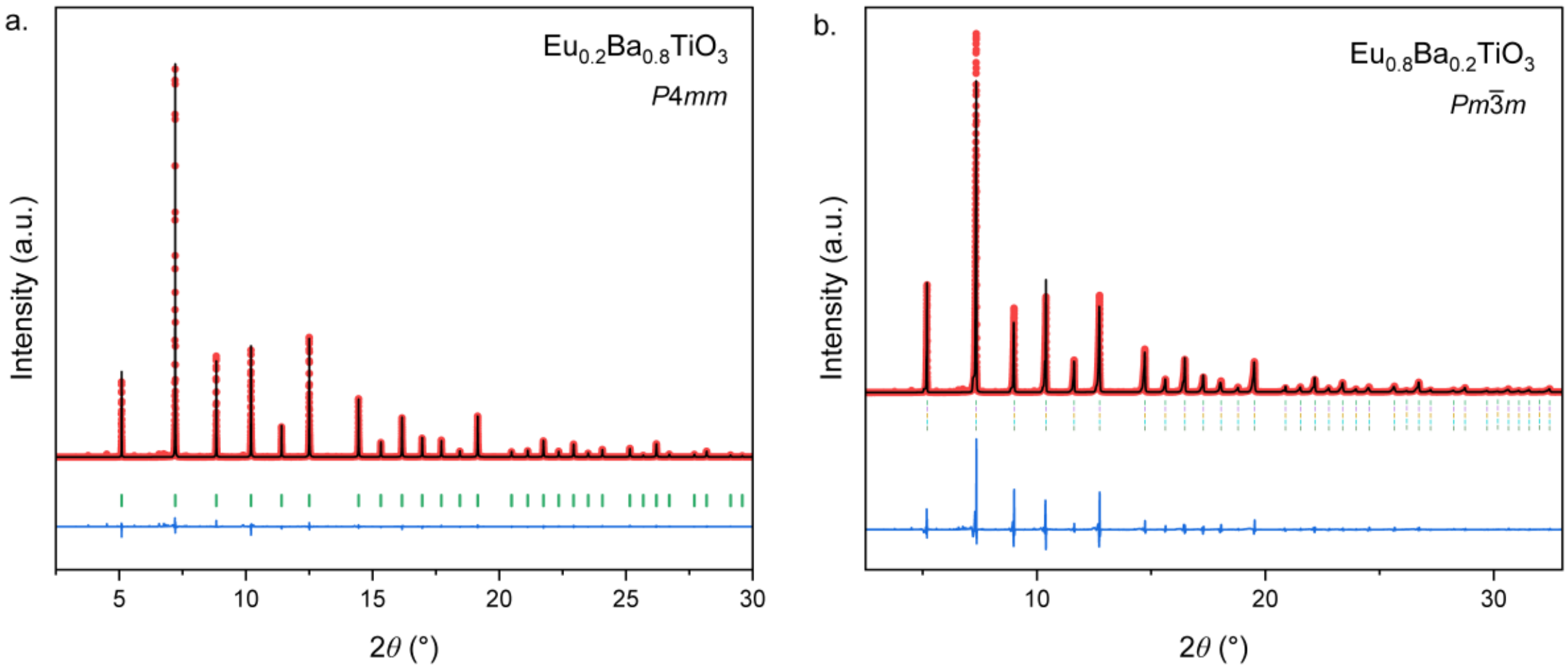

Figure S 5.1. Rietveld refinements of the high-resolution synchrotron XRD data of a. $Eu_{0.2}Ba_{0.8}TiO_3$ in the *P*4*mm* space group, and b. $Eu_{0.8}Ba_{0.2}TiO_3$ in the *Pm*$\bar{3}$*m* space group. The red dots represent the experimental data, the black line the calculated pattern, and the blue line the difference curve. The green hash marks in a. represent the indexed reflections of the *P*4*mm* model, and the different coloured marks in b. represent 5 different cubic models.

Table S 5.5. Structural data from the Rietveld refinements of the high-resolution synchrotron XRD patterns of $Eu_{0.2}Ba_{0.8}TiO_3$ and $Eu_{0.8}Ba_{0.2}TiO_3$. The atomic positions of $Eu_{0.8}Ba_{0.2}TiO_3$ were kept fixed at theoretical values.

| | $Eu_{0.2}Ba_{0.8}TiO_3$ | $Eu_{0.8}Ba_{0.2}TiO_3$ [1] |
|---|---|---|
| Measurement type | HR-XRD | HR-XRD |
| Wavelength $\lambda$ (Å) | 0.354432(9) | 0.354432(9) |
| Measurement range | 0° ≤ $2\theta$ ≤ 47° | 0° ≤ $2\theta$ ≤ 47° |
| Space group | *P*4*mm* | $Pm\bar{3}m$ |
| *a* (Å) | 3.98615(1) | 3.92(1) |
| *b* (Å) | = *a* | = *a* |
| *c* (Å) | 3.98775(1) | = *a* |
| *V* (Å³) | 63.363(1) | 60.4(7) |
| Overall $B_{iso}$ (Å$^2$) | 0.507(5) | -- |
| $B_{iso}$(Eu)/(Ba) | -- | 0.687 |
| $B_{iso}$(Ti)/(O1) | -- | 0.298 |
| $R_p$ (%) | 5.43 | 4.34 |
| $R_{wp}$ (%) | 8.89 | 6.86 |
| $R_{Bragg}$ (%) | 1.96 | -- |
| $\chi^2$ | 30.7 | 2.53 |

[1] The sample is composed of several cubic phases: Phase 1 (23.8(5)%), Phase 2 (37.7(7)%), Phase 3 (31.7(7)%), Phase 4 (4.37(7)%), and Phase 5 (2.43(4)%). For simplification purposes, a weighted average lattice parameter and standard deviation are shown. The average Eu/Ba occupancy is 0.81/0.19. The refinement was performed with the TOPAS6.0 program.

Table S 5.6. Atomic positions data from the Rietveld refinements of the high-resolution synchrotron XRD pattern of $Eu_{0.2}Ba_{0.8}TiO_3$. The atomic positions and occupancy of Ti were kept fixed on the nominal values. O(1) occupies the 1*b* Wyckoff position, and O(2) occupies the 2*c* position.

| | *x* | *y* | *z* | Occupancy (occ) |
|---|---|---|---|---|
| Eu/Ba | 0 | 0 | -0.000(1) | 0.216/0.784(1) |
| Ti | $1/2$ | $1/2$ | $1/2$ | 1 |
| O(1) | $1/2$ | $1/2$ | -0.021(4) | 1.00(2) |
| O(2) | $1/2$ | 0 | 0.464(2) | 1.09(1) |

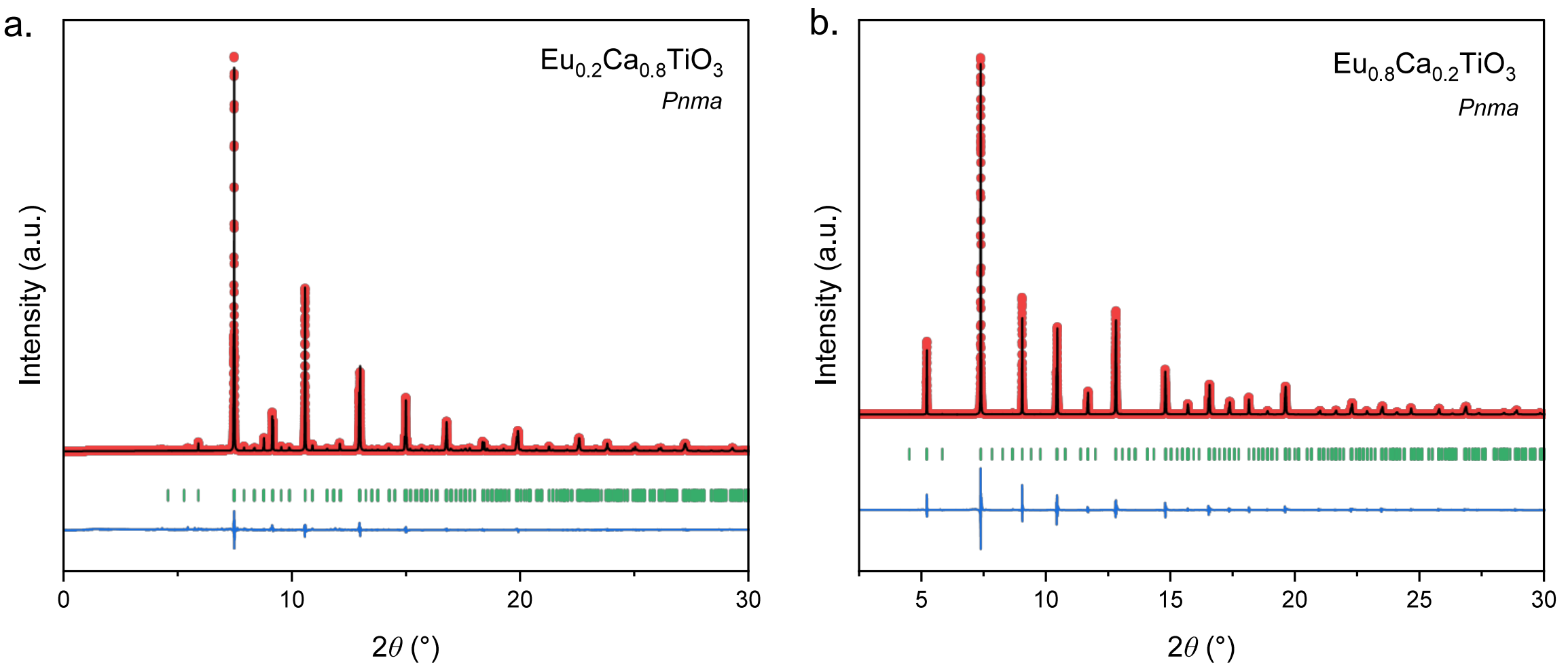

Figure S 5.2. Rietveld refinements of the high-resolution synchrotron XRD data of a. $Eu_{0.2}Ca_{0.8}TiO_3$, and b. $Eu_{0.8}Ca_{0.2}TiO_3$ in the *Pnma* space group. The red dots represent the experimental data, the black line the calculated pattern, and the blue line the difference curve. The green hash marks represent the indexed reflections of the *Pnma* model.

Table S 5.7. Structural data from the Rietveld refinements of the high-resolution synchrotron XRD patterns of $Eu_{0.2}Ca_{0.8}TiO_3$ and $Eu_{0.8}Ca_{0.2}TiO_3$.

| | $Eu_{0.2}Ca_{0.8}TiO_3$ | $Eu_{0.8}Ca_{0.2}TiO_3$ |
|---|---|---|
| Measurement type | HR-XRD | HR-XRD |
| Wavelength $\lambda$ (Å) | 0.354432(9) | 0.354432(9) |
| Measurement range | $0° \leq 2\theta \leq 47°$ | $0° \leq 2\theta \leq 47°$ |
| Space group | *Pnma* | *Pnma* |
| *a* (Å) | 5.45182(1) | 5.50386(8) |
| *b* (Å) | 7.68172(1) | 7.80520(2) |
| *c* (Å) | 5.42061(1) | 5.50277(8) |
| *V* (Å³) | 227.011(1) | 236.392(5) |
| Overall $B_{iso}$ (Å²) | 0.383(4) | 0.477(6) |
| $R_p$ (%) | 7.58 | 11.7 |
| $R_{wp}$ (%) | 10.1 | 16.0 |
| $R_{Bragg}$ (%) | 9.94 | 5.65 |
| $\chi^2$ | 16.4 | 46.0 |

Table S 5.8. Atomic positions data from the Rietveld refinements of the high-resolution synchrotron pattern of $Eu_{0.2}Ca_{0.8}TiO_3$. The atomic positions and occupancy of Ti were kept fixed at nominal values. The occupancies of O(1) and O(2) were fixed at nominal values. O(1) occupies the 4*c* Wyckoff position, and O(2) occupies the 8*d* position.

| | *x* | *y* | *z* | occ |
|---|---|---|---|---|
| Eu/Ca | 0.5233(1) | $^1/_4$ | 0.0051(1) | 0.192/0.808(1) |
| Ti | 0 | 0 | 0 | 1 |
| O(1) | 0.4885(3) | $^1/_4$ | 0.5672(4) | 1.028(4) |
| O(2) | 0.2845(2) | 0.0318(2) | 0.2158(3) | 1.009(2) |

Table S 5.9. Atomic positions data from the Rietveld refinement of the high-resolution synchrotron pattern of $Eu_{0.8}Ca_{0.2}TiO_3$. The atomic positions and occupancy of Ti were kept fixed at nominal values. The occupancies of O(1) and O(2) were fixed at nominal values. O(1) occupies the 4*c* Wyckoff position, and O(2) occupies the 8*d* position.

| | *x* | *y* | *z* | occ |
|---|---|---|---|---|
| Eu/Ca | 0.5033(7) | $^1/_4$ | 0.0023(2) | 0.796/0.204(2) |
| Ti | 0 | 0 | 0 | 1 |
| O(1) | 0.518(7) | $^1/_4$ | 0.539(2) | 1 |
| O(2) | 0.207(7) | 0.011(2) | 0.251(6) | 1 |

### S5.3. 3D Electron Diffraction

For the orthorhombic *Pnma* majority phase of $Eu_{0.2}Ca_{0.8}Ti(O,N)_3$, the *A*-site was refined as a mixed Eu/Ca position with constrained occupancies, giving occ(Eu) = 0.170(8) and occ(Ca) = 0.830(8). For the cubic $P2_13$ minority phase in the same $Eu_{0.2}Ca_{0.8}Ti(O,N)_3$ sample, two crystallographically distinct *A*-sites were refined. The first site shows occ(Eu1) = 0.494(8) and occ(Ca1′) = 0.506((8), while the second site refines to occ(Ca2) = 0.611(12) and occ(Eu2′) = 0.389(12). These occupancies indicate a more Eu-rich *A*-site environment compared to the *Pnma* phase.

All atoms in both refinements were treated with isotropic displacement parameters. Occupancies were constrained such that occ(Eu) + occ(Ca) = 1 for each *A*-site. Nitrogen could not be refined reliably in the 3DED datasets and is therefore omitted from both structural models.

Table S 5.10. 3DED crystallographic parameters of the *Pnma* and *P*$2_1$3 phases of $Eu_{0.2}Ca_{0.8}Ti(O,N)_3$.

| | $Eu_{0.2}Ca_{0.8}Ti(O,N)_3$ | |
|---|---|---|
| Phase | *Pnma* | *P*$2_1$*3* |
| Measurement type | 3D electron diffraction (3DED), dynamical refinements | |
| Radiation/ $\lambda$ (Å) | Electrons, $\lambda$ = 0.02508 | |
| Instrument | Thermo Scientific Talos FEG TEM (200 kV) | |
| Volume (Å) | 213.0(1) | 430.5(2) |
| Calculated density (g cm$^{-3}$) | 4.834 | 5.721 |
| Unit cell content | $Eu_{0.17}Ca_{0.83}TiO_3$ | $Eu_{0.44}Ca_{0.56}TiO_3$ |
| Unique reflections | 2250 | 2917 |
| Observed reflections (I > 2σI) | 2179 | 1603 |
| *a* (Å) | 5.352(1) | 7.551(1) |
| *b* (Å) | 7.523(1) | = *a* |
| *c* (Å) | 5.289(1) | = *a* |
| *R*1 (obs) (%) | 13.80 | 25.99 |
| *wR* (obs) (%) | 16.91 | 25.65 |
| GOF (obs) | 10.53 | 6.82 |

## S6. Electron Energy-Loss Spectroscopy (EELS) and Energy Dispersive Spectroscopy (EDS) of $Eu_{0.2}Ca_{0.8}Ti(O,N)_3$

Elemental distributions of Eu, Ca, Ti, O, and N were mapped by both STEM-EDS and STEM-EELS (Figure S 6.1). For the EELS data, the N-K, O-K, and Ti-L edges were each integrated over a 30 eV window. Because absolute EELS intensities scale with local sample thickness, the N-K and O-K signals were normalised to the Ti-L edge, which served as the internal thickness reference, as Ti is present throughout the structure and exhibits a strong edge signal. Negative N-K integrals indicate nitrogen below the detection limit and were excluded from further quantitative analysis.

Four representative regions were classified by nitrogen content as N-rich, N-medium-rich, N-poor, and N-non-existent, and the corresponding N/Ti and O/Ti ratios were quantified for each (Figure S 6.1). The N-rich region showed the highest N/Ti and lowest O/Ti ratio among the regions analysed, and is co-located with a locally elevated Ti and N signal, consistent with a Ti- and N-enriched local environment resembling TiN.

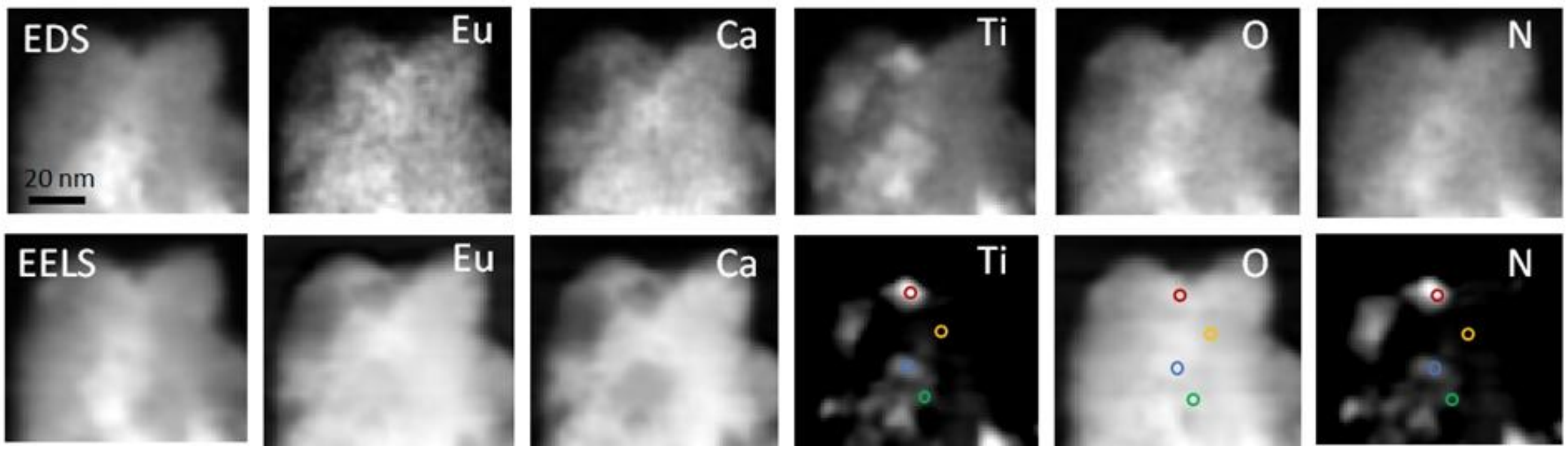

Figure S 6.1. STEM-EDS (top row) and STEM-EELS (bottom row) elemental maps of $Eu_{0.2}Ca_{0.8}Ti(O,N)_3$, showing Eu, Ca, Ti, O, and N distributions. In the EELS maps, coloured points highlight four representative nanoscale regions identified through post-processing and classified by nitrogen content: N-rich (red), N-medium-rich (blue), N-poor (green), and N-non-existent (yellow). The N-rich point shows the highest N/Ti and lowest O/Ti ratio among the regions analysed and is co-located with a locally elevated Ti and N signal, consistent with a Ti- and N-enriched local environment resembling TiN.

## S7. Optical properties – $Eu_{1-x}Ba_xTi(O,N)_3$ / $Eu_{1-x}Ca_xTi(O,N)_3$

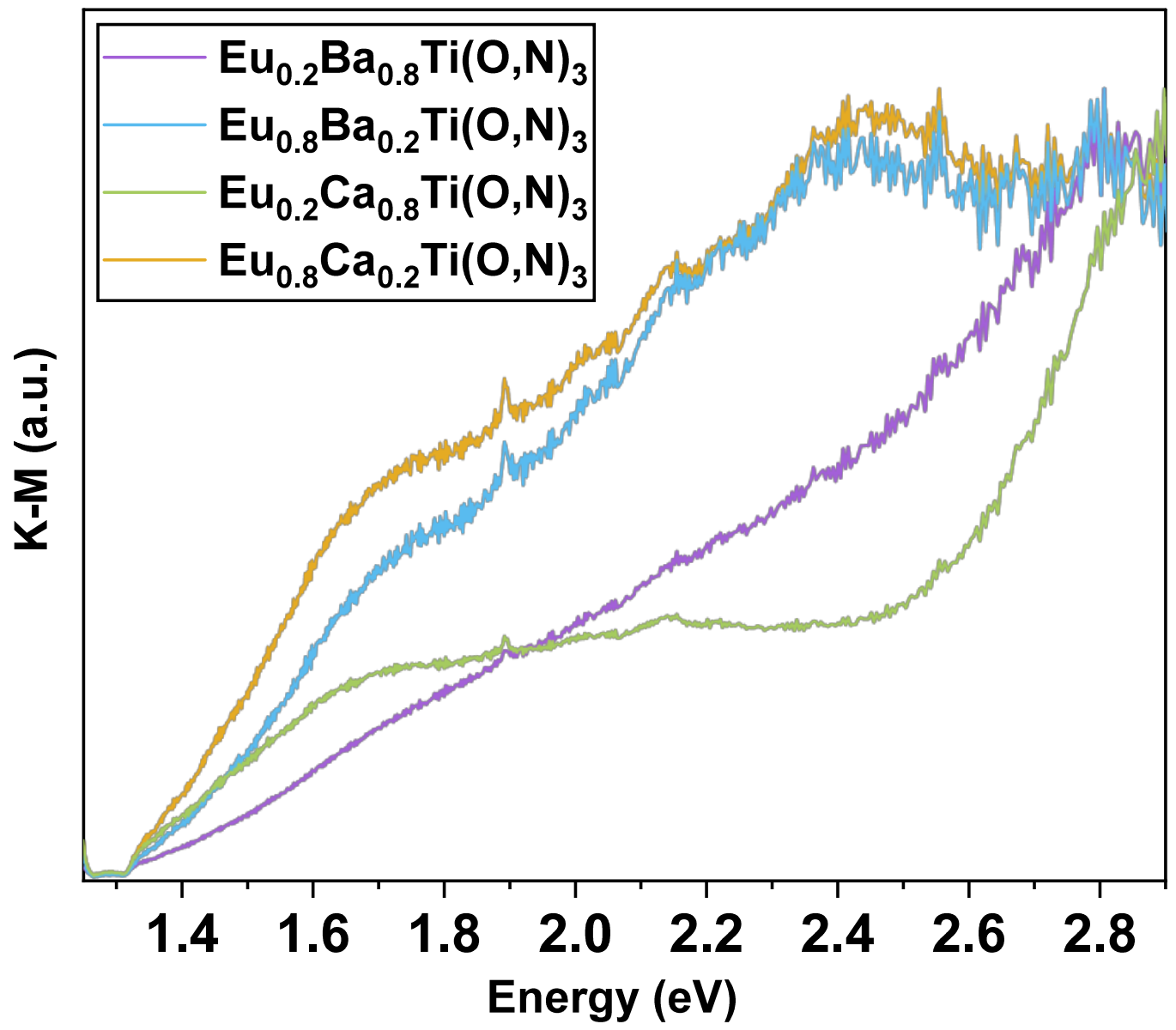

Figure S 7.1. UV-Vis Diffuse Reflectance Spectra, plotted in K-M vs Energy, for $Eu_{1-x}Ba_xTi(O,N)_3$ and $Eu_{1-x}Ca_xTi(O,N)_3$, ($x$ = 0.2, 0.8).

## S8. Electronic structure – $Eu_{1-x}Ba_xTi(O,N)_3$ / $Eu_{1-x}Ca_xTi(O,N)_3$

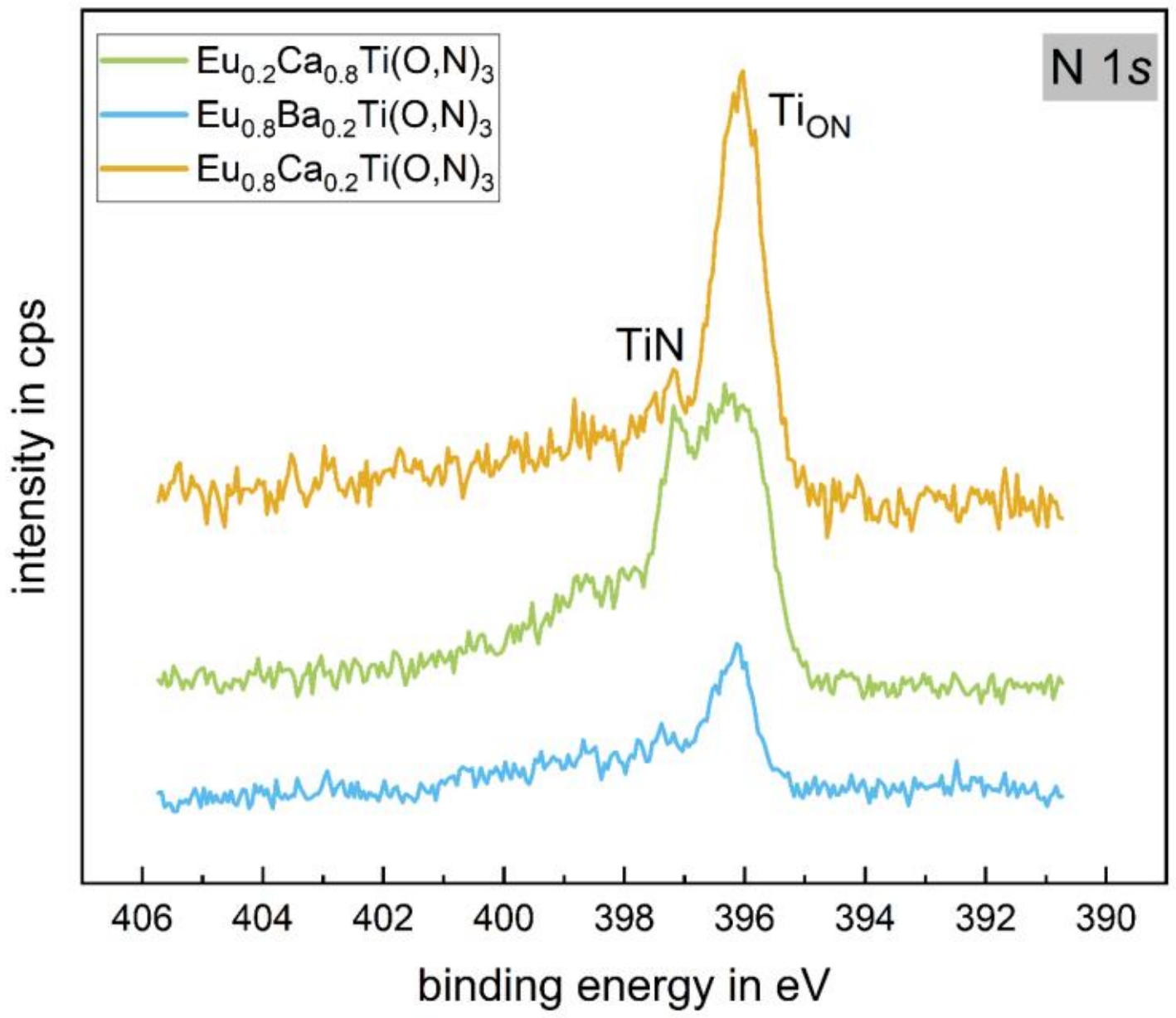

Figure S 8.1. Synchrotron-based X-ray photoelectron spectra of the N 1*s* core level for the perovskite oxynitrides. Measured at SoLiAS@OPUS at BESSY II with an excitation energy of $h\nu$ = 1046 eV ($E_{kin}$ = ~650 eV).

### S8.1. Discussion Ti 2*p* and Ti L-edge and TiN

The Ti $2p_{3/2}$ core level spectra were fit with a Shirley background and deconvoluted using Gaussian-Lorentzian fits (GL(30)) to elucidate the local chemical environment and oxidation states of titanium. For the oxide samples, one main peak was observed, assigned to the $Ti^{4+}$ state, corresponding to the standard tetravalent configuration of the titanium *B*-site cation in the ideal $ABO_3$ perovskite-type lattice [47]. Additionally, a minor component may be present at a lower binding energy, approximately 1.5 eV below the $Ti^{4+}$ main peak. This feature is attributed to a minor contribution of reduced $Ti^{3+}$ species, which is typically associated with oxygen vacancies formed during processing under reducing conditions. However, because the contribution is small, this component was not fitted to avoid overparameterization.

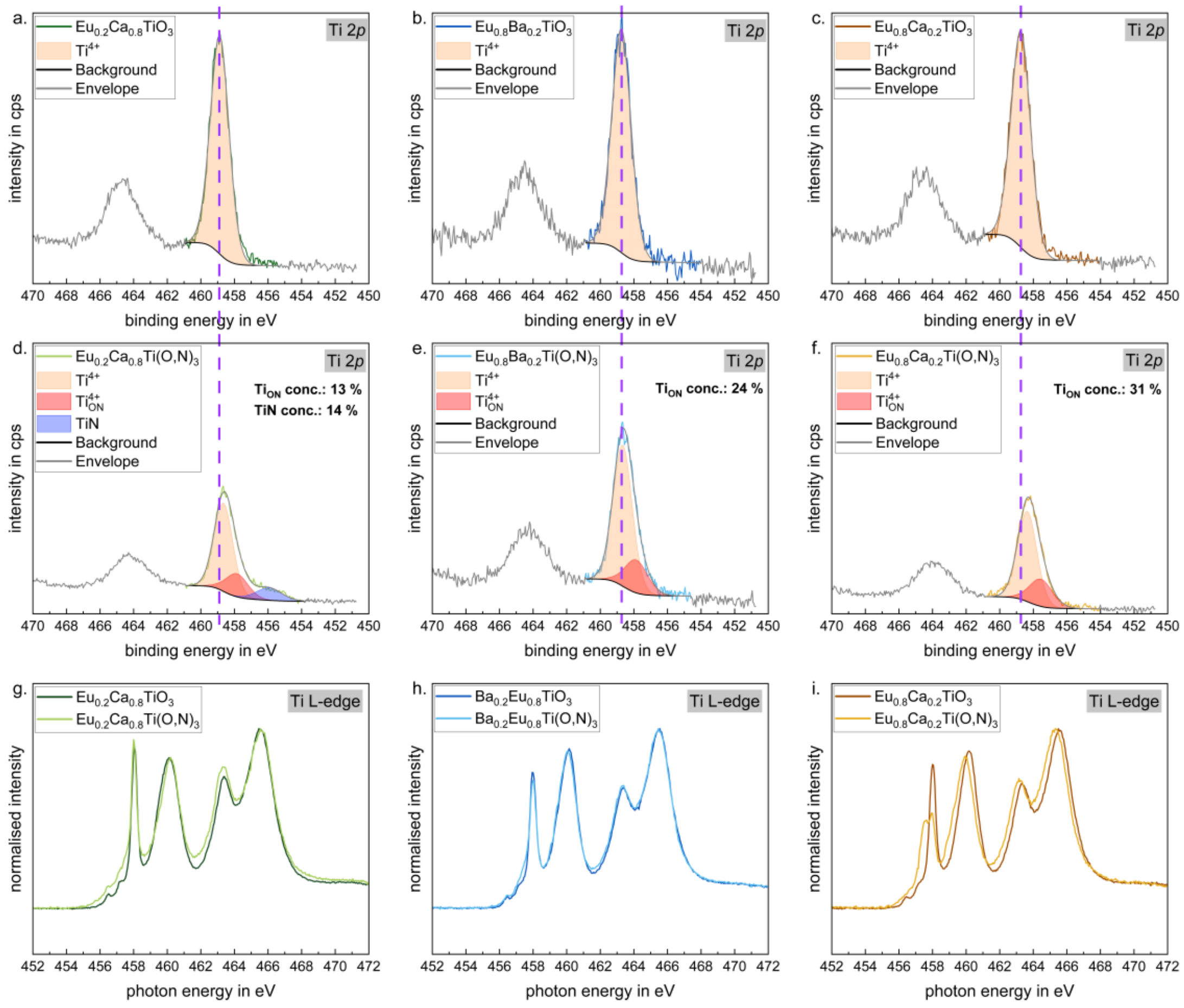

Figure S 8.2. Synchrotron-based X-ray photoelectron spectra of the Ti 2$p_{3/2}$ core level of the a.-c. oxide and d.-f. oxynitride samples, fitted using Gaussian–Lorentzian line shapes. The intensity of each oxide and oxynitride with the same composition is shown in the same scale. In order to compare the binding energy shift a vertical line was drawn from the oxide to the oxynitride (in purple). Measured at SoLiAS@OPUS at BESSY II with an excitation energy of $h\nu$ = 1108 eV ($E_{kin}$ = ~650 eV). g.-i. Near-Edge X-ray Absorption Fine Structure (NEXAFS) spectra recorded at the Ti $L_{2,3}$-edge for the Ca- and Ba-containing samples before and after ammonolysis. NEXAFS was recorded in Auger electron yield (AEY) mode. The NEXAFS data was normalised according to the edge jump.

When the samples were fit with a single peak after ammonolysis, all oxynitrides showed a higher Ti 2$p_{3/2}$ FWHM. This led to the conclusion that another component, likely titanium atoms in a mixed oxynitride coordination environment ($TiO_{6-x}N_x$), is present, which shows the largest impact for the highest nitrogen incorporation in $Eu_{0.8}Ca_{0.2}Ti(O,N)_3$. Due to this, another component was added to the fit, while keeping the FWHM of the oxide fixed. This $Ti^{4+}_{ON}$ component was implemented with a FWHM of 1.2 times that of the oxide and a binding energy shift of 0.8 eV. Since it is assumed that, due to the $Eu^{2+/3+}$ charge transition, oxygen vacancy formation is hindered, little additional $Ti^{3+}$ is expected, which is why it was not taken into consideration in the fit.

However, even after isolating the discrete $Ti^{4+}_{ON}$ coordination environments, the primary oxide-like $Ti^{4+}$ peak of $Eu_{0.8}Ca_{0.2}Ti(O,N)_3$ still exhibits a downward shift in binding energy greater than what can be accounted for by the Fermi level shift alone. While the $Ti^{4+}_{ON}$ peak captures the effects of the first coordination sphere, this overarching shift of the entire lattice network seems to be additionally driven by global changes in covalency. This additional chemical shift is probably observed because nitrogen (3.04) is less electronegative than oxygen (3.44) on the Pauling scale, the replacement of $O^{2-}$ by $N^{3-}$ reduces the overall ionic character of the extended anion network.

For $Eu_{0.2}Ca_{0.8}Ti(O,N)_3$, a third distinct component was required to accurately fit the spectra. This additional feature is located at a significant downward shift of approximately 3.0 eV relative to the $Ti^{4+}$ position, which is characteristic of titanium nitride (TiN) [47,48]. TiN has previously been discussed as forming for higher Ca contents in $Eu_{1-x}Ca_xTi(O,N)_3$ by Widenmeyer et al. [12]. This peak assignment is fully corroborated by the neutron diffraction data refinements and can also clearly be distinguished in the N 1s spectrum in Figure S8.1. This selective phase segregation provides compelling evidence for a saturation-controlled reaction mechanism during high-temperature ammonolysis. Exposure to the reducing and nitriding ammonia atmosphere no longer alters the bulk composition; instead, it induces a localised structural decomposition, driving the exsolution of titanium to form secondary TiN once the sample reaches its limit for nitrogen incorporation, when the $Eu^{2+/3+}$ charge transition is complete.

The analysis of the Ti L-edge NEXAFS of the Ca-containing samples reveals another difference. While the $Eu_{0.8}Ba_{0.2}Ti(O,N)_3$ shows only minor differences to its oxide counterpart, $Eu_{0.2}Ca_{0.8}Ti(O,N)_3$ already shows larger differences around 456 eV to 458 eV photon energy. The $Eu_{0.8}Ca_{0.2}Ti(O,N)_3$ sample shows an even larger difference, even showing a splitting of the $t_{2g}$ feature. This behaviour is well rationalised by the crystal-field multiplet calculations by de Groot et al., which demonstrate that lowering the local coordination symmetry from a perfect octahedron breaks the degeneracy of the $t_{2g}$ subbands, resulting in distinct peak splitting or multi-peak profiles that reflect severe local polyhedral distortions [49].

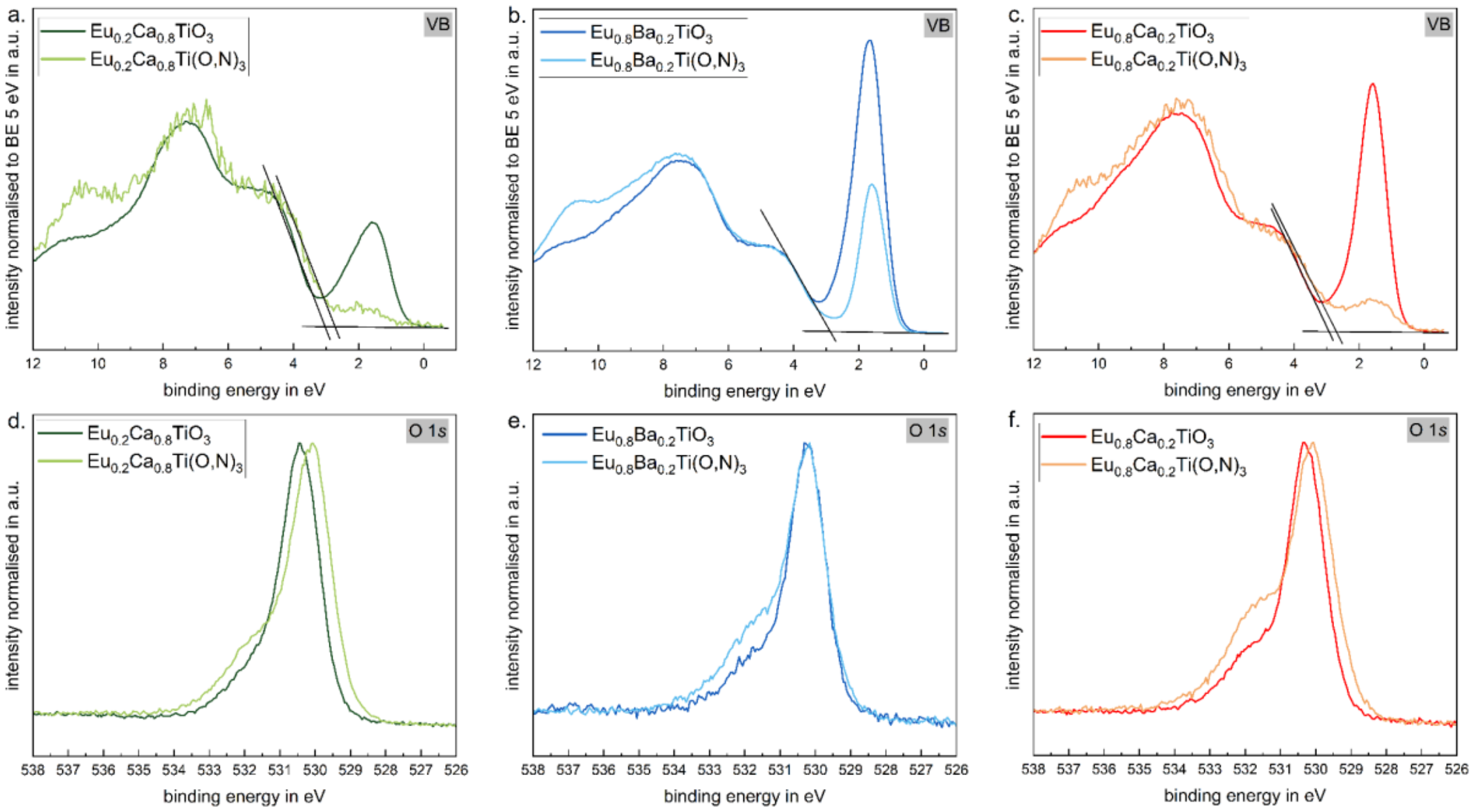

Figure S 8.3. Synchrotron-based X-ray photoelectron spectra of the valence band (a.-c.) and O 1*s* (d.-f.) detail spectra for the perovskite-type oxides and their corresponding ammonolysed oxynitride counterparts. Measured at SoLiAS@OPUS at BESSY II with an excitation energy of $h\nu$ = 650 eV ($E_{kin}$ = ~650 eV), for the VB, and $h\nu$ = 1180 eV ($E_{kin}$ = ~650 eV), for O 1*s*. The spectra were normalised from minimum to maximum to account for intensity differences between the oxide and oxynitride samples.

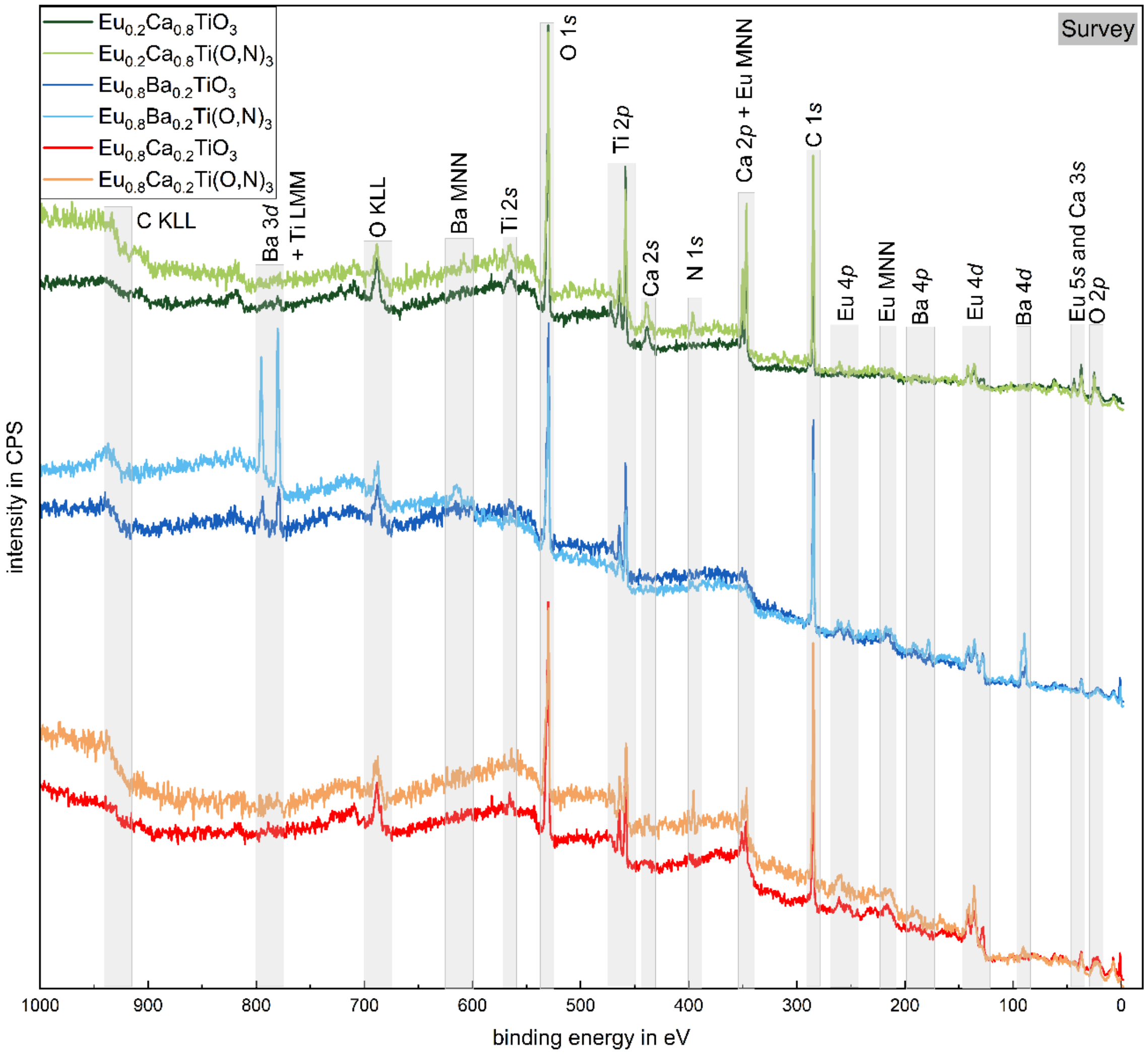

Figure S 8.4. Synchrotron-based X-ray photoelectron survey spectra for the perovskite oxides and their corresponding ammonolysed oxynitride counterparts. Measured at SoLiAS@OPUS at BESSY II with an excitation energy of *hv* = 1200 eV. The plots were shifted along the y-axis for easier comparison.

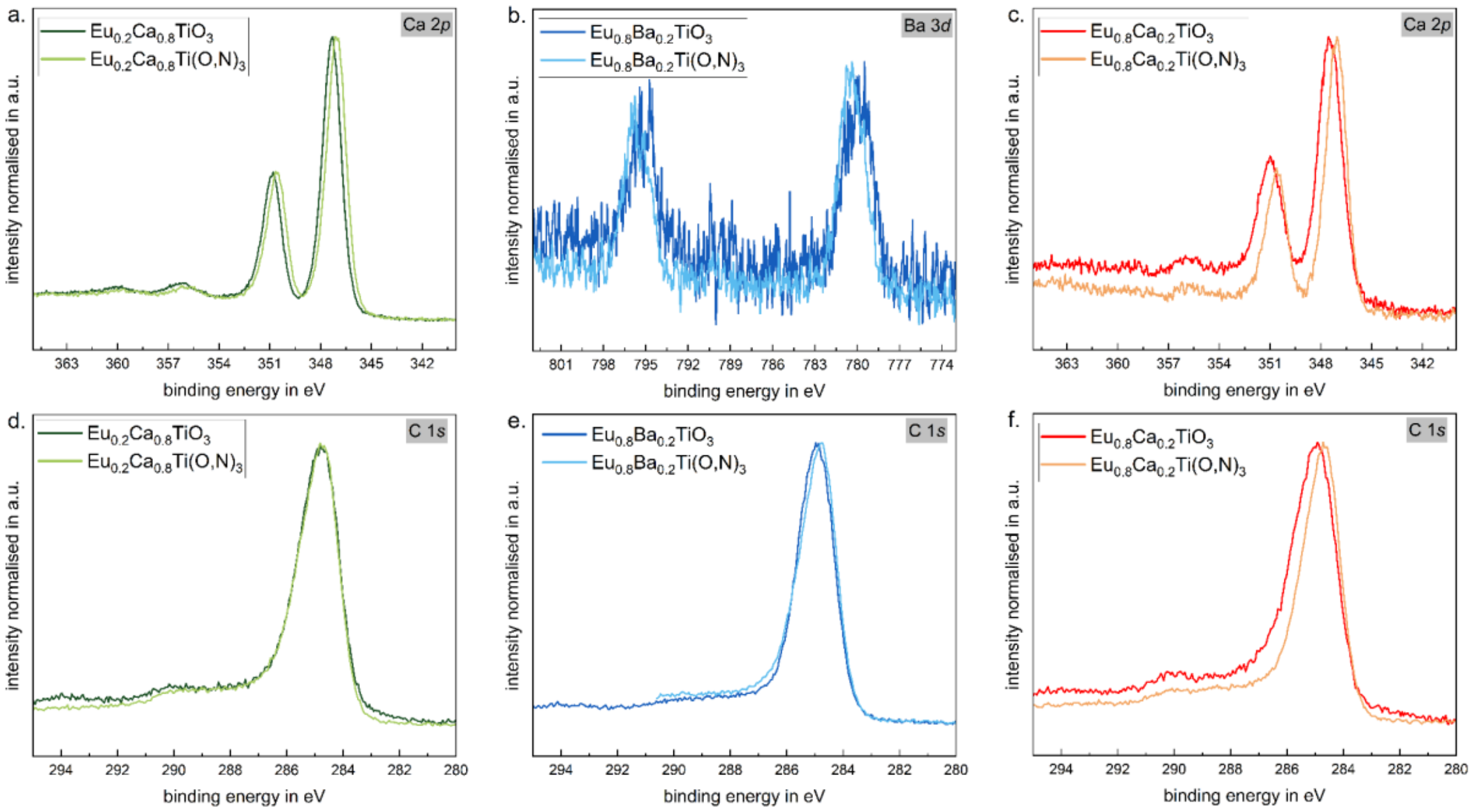

Figure S 8.5. Synchrotron-based X-ray photoelectron spectra of the Ca 2*p* (a. and c.), Ba 3*d* (b.) and C 1*s* (d.-f.) detail spectra for the perovskite-type oxides and their corresponding ammonolysed oxynitride counterparts. Measured at SoLiAS@OPUS at BESSY II with an excitation energy of $hv$ = 997 eV ($E_{kin}$ = ~650 eV) for Ca 2*p*, $hv$ = 1430 eV ($E_{kin}$ = ~650 eV) for Ba 3*d* and $hv$ = 935 eV ($E_{kin}$ = ~650 eV) for C 1*s*. The spectra were normalised from minimum to maximum to account for intensity differences between the oxide and oxynitride samples.

Table S 8.1. Peak positions for the perovskite-type oxide and corresponding ammonolysed oxynitride counterparts, extracted from the synchrotron-based X-ray photoelectron spectra, for O 1*s*, Ti 2*p*, Ca 2*p* or Ba 3*d*, and the valence band maximum. In brackets, the change in binding energy from oxide to oxynitride is shown.

| Binding energy in eV | $Eu_{0.2}Ca_{0.8}TiO_3$ | $Eu_{0.2}Ca_{0.8}Ti(O,N)_3$ | $Eu_{0.8}Ba_{0.2}TiO_3$ | $Eu_{0.8}Ba_{0.2}Ti(O,N)_3$ | $Eu_{0.8}Ca_{0.2}TiO_3$ | $Eu_{0.8}Ca_{0.2}Ti(O,N)_3$ |
|---|---|---|---|---|---|---|
| O 1*s* | 530.38 | 530.07 (-0.31) | 530.19 | 530.18 (-0.01) | 530.23 | 530.02 (-0.21) |
| Ti 2*p* | 458.87 | 458.59 (-0.28) | 458.74 | 458.68 (-0.06) | 458.74 | 458.35 (-0.39) |
| Ca 2*p* | 347.35 | 347.09 (-0.26) | | | 347.40 | 347.15 (-0.25) |
| VBM | 3.05 | 2.75 (-0.30) | 2.90 | 2.90 (0) | 2.95 | 2.70 (-0.25) |

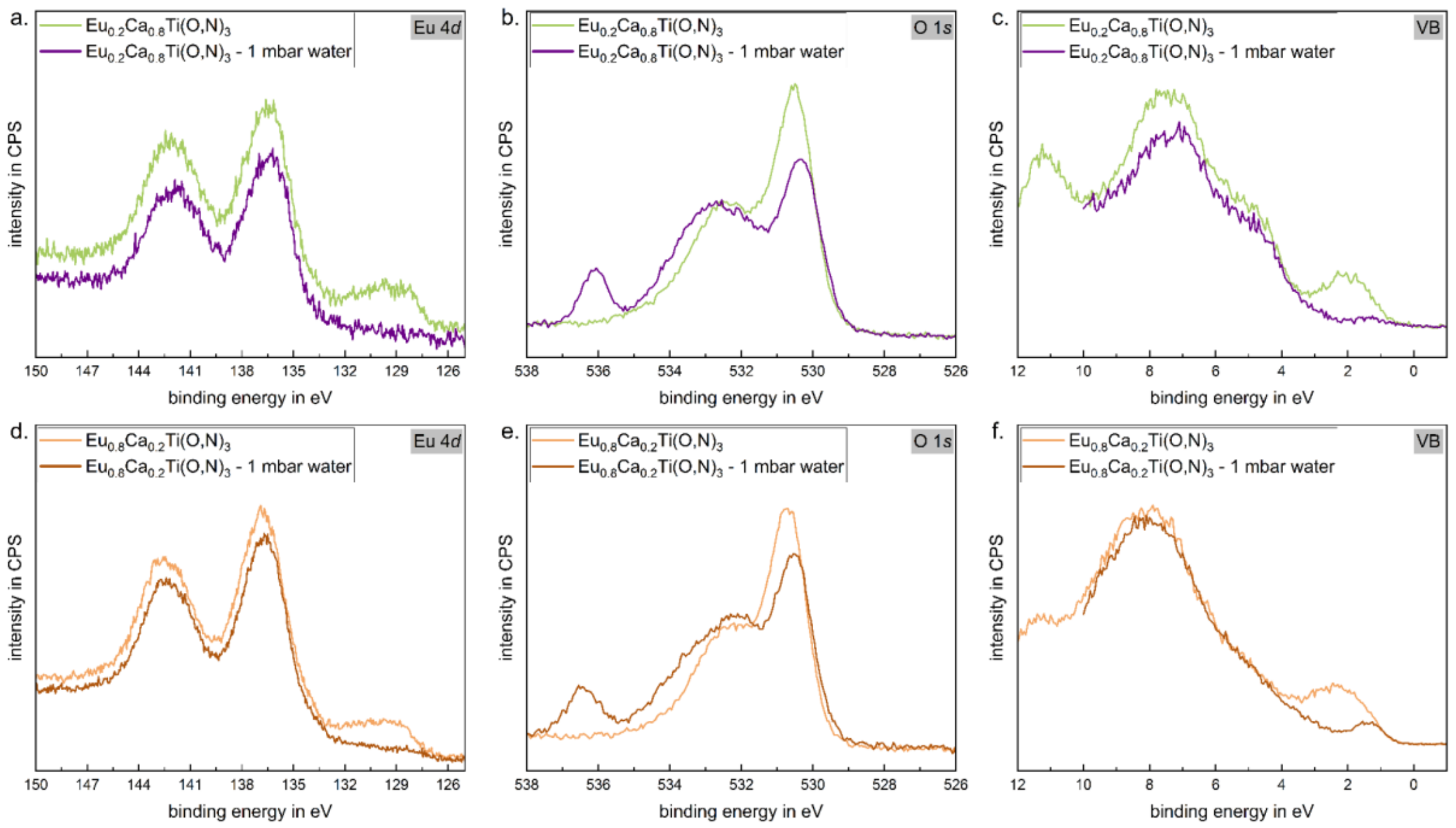

Figure S 8.6. X-ray photoelectron spectra of a.+ d. the Eu 4*d* core levels, b.+ e. O 1*s* core levels and c.+ f. valence band spectra for the oxynitride $Eu_{1-x}Ca_xTi(O,N)_3$ systems. The surfaces were exposed to 1 mbar water vapour at room temperature utilising Near-Ambient pressure XPS (NAP-XPS), measured at TUM with Al $K_\alpha$ radiation.

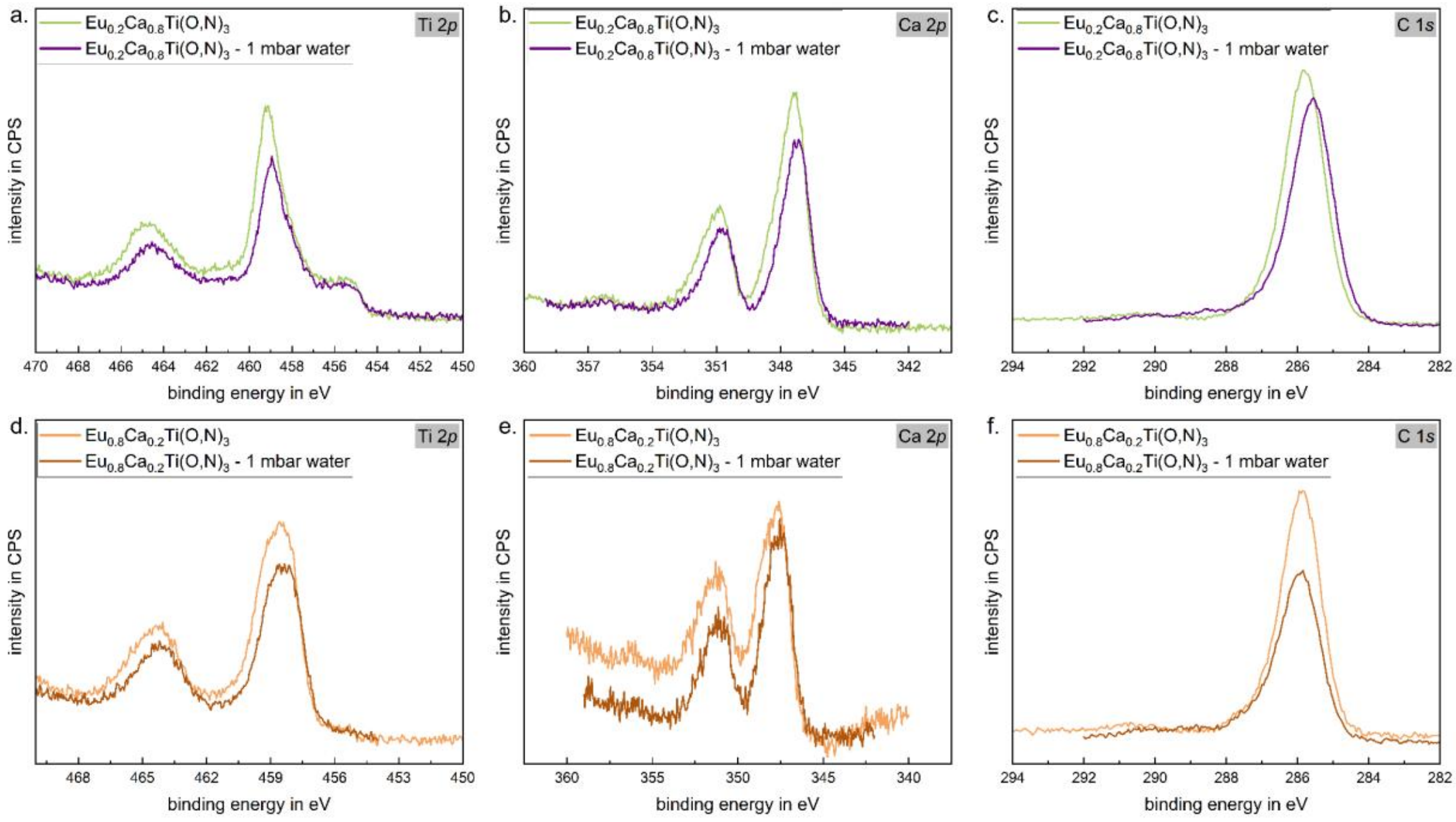

Figure S 8.7. X-ray photoelectron spectra of a.+ d. the Ti 2*p* core levels, b.+ e. Ca 2*p* core levels and c.+ f. *C* 1*s* core level for the oxynitride $Eu_{1-x}Ca_xTi(O,N)_3$ systems. The surfaces were exposed to 1 mbar water vapour at room temperature utilising Near-Ambient pressure XPS (NAP-XPS), measured at TUM with Al Kα radiation.